\documentstyle[aps,twocolumn,epsfig]{revtex}
\begin{document}
\draft
\title{\large \bf RELATIVISTIC INSTANT--FORM APPROACH TO
THE STRUCTURE OF
TWO-BODY COMPOSITE SYSTEMS}

\author{A.F.Krutov\thanks{Electronic address:
krutov@ssu.samara.ru}}

\address {\it Samara  State University, 443011, Samara, Russia}

\author{V.E.Troitsky\thanks{Electronic address:
troitsky@theory.npi.msu.su}}
\address {\it Nuclear Physics
Institute, Moscow State University, 119899, Moscow, Russia}

\date{January, 2001}
\maketitle

\begin{abstract}
A new approach to the electroweak properties of
two--particle composite
systems is developed. The approach is based on
the use of the instant form of relativistic Hamiltonian
dynamics.
The main novel feature of this
approach is the new method of construction of the matrix element
of the electroweak current operator.
The electroweak current matrix element satisfies the
relativistic covariance conditions and in the case of the electromagnetic
current also the conservation law automatically.
The properties of the system as well as the
approximations are formulated in terms of form factors.
The approach makes it possible to formulate relativistic
impulse approximation 
in such  a way that the Lorentz--covariance of the current
is ensured. In the electromagnetic case the current conservation
law is ensured, too.
The results of the calculations are
unambiguous:  they do not depend on the choice of the coordinate
frame and on the choice of "good" components of the current as
it takes place in the standard form of light--front dynamics.
Our approach gives good
results for the pion electromagnetic form factor in the whole
range of momentum transfers available for experiments at present
time, as well as for lepton decay constant of pion.

\end{abstract}
\narrowtext

\section{Introduction.}
All atoms, nuclei, and the main part of the so called
elementary particles are composite systems. That is why the
constructing of correct quantitative methods of calculation
for
composite--particle structure is an important line of
investigations in particle physics. In nonrelativistic dynamics
there exist different correct methods which use model or
phenomenological interaction potentials. However, in the case of
high energy one needs to develop relativistic methods. It is
worth to note that now the experiments on accelerators, in
particular at JLab are performed with such an accuracy that the
treatment of traditionally "nonrelativistic" systems (e.g. the
deuteron) needs to take into account relativistic effects.
Relativistic effects are important also in the treatment of
composite systems of light quarks. In this case the relativistic
effects are significant even at low energy. However, the
relativistic treatment of hadron composite systems is a rather
complicated problem. To solve it one needs, in fact, to solve a
many--particle relativistic problem with, in addition, not
precisely known interaction. Let us note that the use of the
methods of the field theory in this case encounters serious
difficulties. For example, it is well known that the
perturbative QCD can not be used in the case of quark bound
states (see, e.g.,~\cite{Gro93,Kei94W}).

In the present paper we will use the relativistic
constituent model which describes the hadron properties at quark
level in terms of degrees of freedom of constituent quarks. The
constituent quarks are considered as extended objects , the
internal characteristics of which (MSR, anomalous magnetic moments, form
factors) are parameters of model.
As relativistic
variant of constituent model we
choose the method of relativistic Hamiltonian dynamics (RHD)
(see, e.g.,
~\cite{LeS78,KeP91,Coe92,Kli98} and the references therein).

Our aim is to construct a relativistic invariant approach to
electroweak structure of two--particle composite systems. The
main problem  here is the construction of the current operators
\cite{GrR87,ChC88,CoR94,Lev95,LeP98}.

It seems to us that RHD is the method the most adequate for our
purpose. The use of RHD enables one to separate the main degrees
of freedom and so to construct convenient models.

We use one of the forms of RHD, namely the version of the
instant form (IF).

Our approach has a number of features that distinguish it from
other forms of dynamics and other approaches in the frames of
IF.

\begin{itemize}

\item
The electroweak current matrix element satisfies the
relativistic covariance conditions and in the case of the electromagnetic
current also the conservation law automatically.

\item
We propose a modified impulse
approximation (MIA). It is constructed in relativistic invariant
way. This means that our MIA does not depend on the choice of
the coordinate frame, and this contrasts principally with the
"frame--dependent" impulse approximation usually used in instant
form (IF) of dynamics.
\footnote{
It is known that correct impulse approximation (IA) realization
in the frame of traditional version of IF dynamics encounters
difficulties: the standard IA depends on the choice of
the coordinate frame. We show below that IA can be formulated in
an invariant way, the composite system form factors being
defined by the one--particle currents alone.}

\item
Our approach provides with correct and natural
nonrelativistic limit ("the correspondence principle" is
fulfilled).

\item
For composite systems (including the spin 1 case) the
approach guarantees the uniqueness of the solution for form factors
and it does not use such concepts as "good" and "bad" current components.

\item
The approach describes correctly the spin Wigner rotation
and this fact makes it possible to obtain
the correct (QCD) asymptotic.

\end{itemize}

The RHD method as the relativistic theory of composite systems
is based on the direct realization of the Poincar\'e algebra on
the set of dynamical observables on the Hilbert space (see, e.g.,
~\cite{LeS78,KeP91,Coe92,Kli98} and the references therein).
RHD theory of particles lie
between local field theoretic models and nonrelativistic quantum
mechanical models.

Let us describe briefly the main statements of RHD.

As it is known (see, e.g., \cite{Nov75}),
the relativistic invariance of a theory means that there exists
(on the Hilbert space of states) the unitary representation of
the inhomogeneous group
$SL(2,C)$, which is the universal covering of the Poincar\'e
group. The relativistic invariance means the validity of the
Poincar\'e algebra commutation relations for the generators of
space--time translations
$\hat P^\mu$ and rotations $\hat M^{\mu \nu}$.
To construct the representation $SL(2,C)$ means to obtain these
generators in terms of dynamical variables of the system. In the
case of free particle system this program is not difficult to be
realized and the generators
$\hat P^\mu\;,$$\;\hat M^{\mu \nu}$
have clear physical meaning:
$\hat P^0 \equiv \hat H$ --- is the total energy operator,
$\hat {\vec P} = (\hat P^1,\hat P^2,\hat P^3)$ --- is the operator of the
total 3--momentum,
$\hat {\vec J} = (\hat M^{23},\hat M^{31},\hat M^{12})$ ---
the operator of the total angular momentum, $\hat {\vec N} =
(\hat M^{01},\hat M^{02},\hat M^{03})$ ---
the generators of Lorentz boosts.

In the case of interacting systems the situation is different
and the construction of the generators in terms of dynamical
variables encounters some difficulties. To make clear the sense
of these difficulties let us compare RHD with the
nonrelativistic quantum mechanics. In nonrelativistic case it is
the Galilean group that is the group of invariance. In the case
of nonrelativistic interacting system the interaction operator
enters (in additive way) only the generator of time translations
(energy operator). The interaction operator satisfies the usual
conditions of invariance under translations and rotations  and of
independence on the choice of inertial coordinate frame. Under
these conditions the algebraic relations for the generators of
the Galilei group written in terms of dynamical variables remain
to be valid after the interaction including. This is the meaning
of the Galilean invariance of the theory. So, in
nonrelativistic theory of interacting particles it is only one
generator of Galilei group --- the Hamiltonian --- that contains
the interaction. Other generators have the same form as in the
case of free particles. This way of interaction including in the
observables algebra is unique in nonrelativistic case and gives
the unique nonrelativistic dynamics --- the dynamics governed by
the Schr\"odinger equation.

The situation in the relativistic case is quite different. The
structure of the Poincar\'e algebra is of such kind that the
interaction including in the total energy operator alone results
in the breaking of algebraic structure, i.e. in the relativistic
invariance breaking. This is the consequence of the fact that
the Lorentz transformations mix space and time. To preserve the
algebra it is necessary to include the interaction in the other
generators, too. Now the generators of Poincar\'e algebra can be
divided into two classes: containing interaction
(interacting generators) -- hamiltonians, and without
interaction (non-interacting generators).
The latter present the so called kinematical subgroup. This
division, that is the interaction inclusion in the Poincar\'e
group, is not unique. The different ways of such a division
preserving the Poincar\'e algebra result in different types of
relativistic dynamics (see, e.g., \cite{KeP91}).

The idea of this approach --- RHD --- is originated by Dirac.
In ~\cite{Dir49} he considered different ways of description of
the evolution of classical relativistic systems --- different
forms of dynamics. Dirac separated the concept of time as a
coordinate and that of time as a parameter defining the system
evolution.  Consequently, he defined three main forms of
dynamics with different evolution parameters: point (PF),
instant (IF) and light--front (FF) dynamics.  Each of these
forms has its relative advantages and disadvantages. The total
number of possible dynamics is actually five ~\cite{LeS78} so that the
unique nonrelativistic Hamilton description is changed in
relativistic case for, generally, five possibilities. Each of
these possible dynamics is connected with a three--dimensional
hypersurface in the four dimensional space.  The initial
conditions are given on these hypersurfaces and the evolution of
the hypersurfaces is described. The kinematical subgroups are
the invariance groups of the hypersurfaces. In particular, these
hypersurfaces are: hyperboloid $x^\mu\,x_\mu = a^2,\quad
t\;>\;0$ (PF), hyperplane  $t = 0$ (IF), light--cone surface
$x^0 + x^3 = 0$ (FF).

It is worth to notice that RHD and the field theory are quite
different approaches.
The establishment of the
connection between RHD and field theory is a difficult and as
yet unresolved problem.
Contrary to field theory, RHD is dealing
with finite number of degrees of freedom from the very
beginning. This is certainly a kind of a model approach. The
preserving of the Poincar\'e algebra ensures the relativistic
invariance (see for details the Sec.II).
So, the covariance of the description in the frame of RHD is due
to the existence of the unique unitary representation of
the inhomogeneous group $SL(2,C)$ on the Hilbert space of
composite system states with finite number of degrees of
freedom. The success of composite models shows the validity of
approximate relativistic invariant description with fixed number
of particles and a finite number of degrees of freedom.
The similar situation (as was pointed out in ~\cite{LeS78})
takes place in the solid state theory: many of solid state
properties are connected directly with its symmetry group and
the actual form of the particle interaction plays less important
role. RHD is based on the simultaneous action of two fundamental
principles: relativistic invariance and Hamiltonian principle
--- and presents the tool the most adequate to treat the systems
with finite number of degrees of freedom. It is worth to notice
that the mathematics of RHD is similar to that of
nonrelativistic quantum mechanics and permits to assimilate
the sophisticated methods of phenomenological potentials and can
be generalized to describe three or more particles.

Now the problem of the choice of the actual form of RHD arises.

Some time ago it was proved that $S$--matrices are equivalent in
the different dynamics forms ~\cite{SoS78}. This fact is
interesting but it does not mean the absolute  equivalence of the
forms.  First, there are problems which can not be reduced to
$S$--matrix, e.g.,  the calculation of form factors. Second, one has
to keep in mind that any concrete calculation uses some
approximations; the approximations usually used in different
forms of dynamics are nonequivalent.

Our point of view is the following.
One must choose the form of dynamics adequate to the problem
in question and to the approximations to be done. It seems us
that this is in the spirit of RHD -- the choosing of the
adequate degrees of freedom.

RHD is widely used in the theory of electromagnetic properties
of composite quark  and nucleon systems
~\cite{Kli98},~\cite{ChC88},~\cite{LeP98},
~\cite{GrK84,Jau91,KrT93,Sch94,CaG96,BaK96,YaO96,Kru97,AlK98,KrT99,BaK00,AlK00,And00,KrS00}.
It was shown that RHD not only presents an interesting
relativistic model but is a fruitful tool and can
compete with other approaches in describing the existing
experimental data ( especially at small and moderate momentum
transfers).
At present time the FF dynamics
is the most developed and used for composite systems
\cite{ChC88,LeP98,GrK84,Jau91,Sch94,CaG96}.
In fact, the light front dynamics has obvious
advantages: a) the minimal number of operators containing the
interaction (three); b) the simple relativistic invariant
separation of the variables into classes of "internal" and
"external" variables (while considering the approach as Hamiltonian
theory with fixed particle number); c) the simple vacuum
structure for the light--front perturbative field theory.
However, there are some difficulties in the FF RHD approach.
In particular, it was shown
~\cite{GrK84,Kei94} that the calculated electromagnetic form
factors for the systems with the total angular momentum
$J = 1$ (the deuteron, the $\rho$ -- meson) vary significantly
with the rotation of the coordinate frame. This ambiguity is
caused by the breaking of the so called angle condition
~\cite{GrK84,Kei94},
that is really by the breaking of the rotation invariance of the
theory. Some of the difficulties of FF dynamics are discussed in
~\cite{Fud90}. A possible way to solve the problem by adding
some new (nonphysical) form factors to the electromagnetic
current was proposed (see ~\cite{CaD98} and references therein).

A different approach to the problem was proposed recently in
Ref.~\cite{LeP98}, where a new method of construction of
electromagnetic current operators in the frame of FF dynamics
was given. The method of
~\cite{LeP98} gives unambiguous deuteron form factors. However,
as the authors of ~\cite{LeP98} note themselves, their
current operator and the one used in Ref.
~\cite{ChC88} are different, since both of them are obtained
from the free one, but in different reference frames, related by
an interaction dependent rotation.

All these facts naturally cause authors to consider other forms
of relativistic Dirac dynamics.

Recently the PF dynamics was considered in the papers
~\cite{Kli98,AlK98,AlK00,And00}. The authors used PF dynamics to
calculate the processes under investigations in JLab
experiments. It is worth to notice that these experiments
enlarge the interest to the RHD approach -- the relativistic
theory which can be used in the region of soft processes.

Now we present a relativistic treatment of the
problem of soft electroweak structure in the framework of
another form -- IF of RHD. IF of relativistic dynamics, although
not widely used, has some advantages.
The calculations can be performed in a natural straightforward
way without special coordinates. IF is particularly convenient
to discuss the nonrelativistic limit of relativistic results.
This approach is obviously rotational invariant, so IF is the
most suitable for spin problems.

We describe the dynamics of composite systems (the constituent
interaction) in the frame of general RHD axiomatics. However,
our approach differs from the traditional RHD by the way of
constructing of matrix elements of local operators. In
particular, our method of description of the electromagnetic
structure of composite systems permits the construction of
current matrix elements satisfying the Lorentz--covariance
condition and the current conservation law.

To construct the current operator in the frame of IF RHD we use
the general method of relativistic invariant parameterization of
matrix elements of local operators proposed as long ago as in
1963 by Cheshkov and Shirokov~\cite{ChS63}.

The method of ~\cite{ChS63} gives matrix elements of the
operators of arbitrary tensor dimension (Lorentz--scalar,
Lorentz-vector, Lorentz--tensor) in terms of a finite number of
relativistic invariant functions -- form factors. The form
factors contain all the dynamical information on the transitions
defined by the operator. That is why a system can be described
in terms of form factors.

The method of parameterization is similar in spirit to the method
of presentation of matrix elements of irreducible tensor
operators on the rotation group in terms of reduced matrix
elements. This method extracts from the matrix element of a
tensor operator a part defining symmetry properties and
selection rules following the well known Wigner--Eckart theorem.

In the review \cite{KeP91}
two possible variants of such kind of representation of matrix
elements in terms of form factors are presented -- the
elementary--particle parameterization and the multipole
parameterization. The variant of parameterization given in
\cite{ChS63} is an alternative one. In \cite{ChS63} the authors
propose the construction of matrix elements in canonical basis
so it can be called canonical parameterization. This method was
developed for the case of composite systems in
~\cite{TrS69,KoT72}. The composite--system form factors in this
approach are in general case the distributions (generalized
functions), they are defined by continuous linear functionals on
a space of test functions.  Thus, for example, the current
matrix elements for composite systems are functionals, generated
by some Lorentz--covariant distributions, and the form factors
are functionals generated by regular Lorentz--invariant
generalized functions. We demonstrate these facts below, in
Sec.III, using a simple model as an example.

It is worth to notice that the statement that the form factors
of a composite system are generalized functions is not something
exotic. This fact takes place in the standard nonrelativistic
potential theory, too (see Sec.III(F)).

The use of canonical parameterization permits to describe the
electroweak properties of composite systems satisfying the
Lorentz--covariance condition on each stage of calculation and
to satisfy the electromagnetic current conservation law when
describing the electromagnetic properties. In our formalism it
is necessary to formulate the composite model features in
slightly unusual way in terms of matrix elements which are
generalized functions.

In particular, the relativistic impulse approximation (IA) has
to be reformulated.

Let us remind the physics of IA.
In IA a test particle
interacts mainly with each component
separately, that is the electromagnetic current of the composite
system can be described in terms of one--particle currents. In
fact, the composite--system current is approximated by the
corresponding free--system current. This means that  exchange
currents are neglected, or, in other words, that there is no
three--particle forces in the interaction of a test particle
with constituents. It is well known that the traditional IA
breaks the Lorentz--covariance of the composite--system current
and the conservation law for the electromagnetic current (see,
e.g., \cite{KeP91} for details).
As we show in the Sec.III(C) one can overcome these difficulties
if one formulates IA in terms of form factors.

It is worth to notice that all known approaches (including the
perturbative quantum field theory (QFT)) encounter difficulties
while constructing a composite--system current operator
satisfying Lorentz--covariance and conservation conditions.
This problem is now discussed widely in the literature
\cite{GrR87,ChC88,CoR94,Lev95,LeP98}.
To satisfy the conservation law in the frame of Bethe--Salpeter
equation and quasipotential equations, for example, it is
necessary to go beyond IA: one has to add the so called
two--particle currents to the current operator. In the case of nucleon
composite systems these
currents are interpreted as meson exchange currents
\cite{CoR94}.  In the case of deuteron this means the
simultaneous interaction of virtual
$\gamma$-- quanta with proton and neutron. However, in Ref.
\cite{VaD95} it is shown that the current conservation law can
be satisfied without such processes, although they contribute to
the deuteron form factor. It seems that at the present time
there is an intention to formulate IA with transformed
conservation properties without dynamical contribution of
exchange currents
\cite{LeP98,AlK00,CaD98}.

Our formalism also gives, in fact, the description of the
covariance properties of the operators in terms of
many--particle as well as one--particle currents. However, the
important feature of our formalism is the fact that form factors
or reduced matrix elements describing the dynamics of
transitions contain only the contributions of one--particle
currents.

So, our approach to the construction of the current operator
includes the following main points:

\noindent
1. We extract from the current matrix element of composite
system the reduced matrix elements (form factors) containing the
dynamical information on the process. In general these form
factors are generalized functions.

\noindent
2. Along with form factors we extract from the matrix element a
part which defines the symmetry properties of the current: the
transformation properties under Lorentz transformation, discrete
symmetries, conservation laws etc.

\noindent
3. The physical approximations which are used to calculate the
current are formulated not in terms of operators but in terms of
form factors.

In this paper we present the main points of
our approach. To make it transparent we consider here only
simple systems with zero total angular momenta, so that
technical details do not mask the essence of the method. We
demonstrate the effectiveness of the approach by calculating the
pion electroweak properties. In this case the canonical
parameterization is very simple and can be realized without
difficulties. The case of more complicated systems needs using a
rather sophisticated mathematics for canonical parameterization
of local operator matrix elements and will be considered
elsewhere.

The paper is organized as follows.
In Sect. II we remind briefly the basic statements of RHD,
especially of IF RHD. Two bases in the state space of composite
system are considered: the basis with individual spins and
momenta and the basis with separated center--of--mass motion.
The Clebsh--Gordan decomposition for the Poincar\'e group
which connects these bases is given. The IF wave functions of
composite systems are defined. In Sect. III our approach to
relativistic theory of two--particle composite systems and their
electroweak properties is presented. A simple model is
considered in details: two spinless particles in the $S$-state
of relative motion, one of the particles being uncharged. The
electromagnetic form factor of the system is derived. The
standard conditions for the current operator are discussed. The
modified impulse approximation (MIA) is proposed. The results of
IA and MIA are compared. The nonrelativistic limit is considered.
The connection between the presented
version of IF RHD and the dispersion relations in mass is
established. In Sect.IV
the developed formalism is used in the case of the system of two
particles with spins 1/2. The pion electromagnetic form factor
and the lepton decay constant are derived. The model parameters
are discussed and the comparison of the results with the
experimental data is given. The results of calculations in IA
and MIA are compared and are shown to differ significantly. In
Sect.V the conclusion is given.

\section{Relativistic Hamiltonian Dynamics.}

In this Section some basic equations of RHD and some relations
from relativistic spin theory are briefly reviewed.

The relativistic invariance of a theory means that the unitary
representation of the Poincar\'e group is realized on the
Hilbert space of system states (see, e.g., ~\cite{Nov75}).
In this case the structure of the Poincar\'e algebra can be
defined on the set of observables $(\hat M^{\mu \nu }, \hat P^\sigma)$ 
$$
[\hat M^{\mu \nu }, \hat P^\sigma ] =
-i(g^{\mu \sigma }\hat P^\nu - g^{\nu  \sigma }\hat P^\mu ),
$$
$$
[\hat M^{\mu \nu },\hat M^{\sigma\rho }] =
-i(g^{\mu \sigma }\hat M^{\nu \rho } - g^{\nu \sigma}\hat M^{\mu \rho })
- (\sigma \leftrightarrow \rho ),
$$
\begin{equation}
[\hat P^\mu, \hat P^\nu ] = 0.
\label{algebra}
\end{equation}

In (\ref{algebra}) $g^{\mu \nu}$ is the metric tensor in
Minkowski space. As we have mentioned, in the case of the
particles with interaction the realization of the Poincar\'e
algebra on the set of observables is more complicated than in
the  case of the free particles. Let us
consider one key
commutator from the set of commutators (\ref{algebra})
(see, e.g., \cite{Kei94W}):
\begin{equation}
[\hat P^j \hat N^k ] =
i\,\delta^{jk}\,\hat H\;
\label{key}
\end{equation}
(we use notations given in the Introduction; $j,k=1,2,3$).

Since $\hat H$ is interaction dependent for non-trivial systems,
either $\hat{\vec P}\;,\;\hat{\vec N}$, or some combination of
$\hat{\vec P}$ and $\hat{\vec N}$
also must be interacting. To preserve the commutation relations
(\ref{algebra}) one has to make other generators depending
on the interaction, too. So, the generators occur to fall
into two groups: the generators which are independent of the
interaction and form the so called kinematical subgroup, and
the generators depending on the interaction -- Hamiltonians.
This division is not unique.
Different ways to obtain kinematical subgroups
result in different forms of dynamics.

In this paper we use the so called instant form dynamics
(IF). In this form the kinematical subgroup contains the
generators of the group of rotations and translations in
the three--dimensional Euclidean space:
\begin{equation}
 \hat{\vec J}\>,\quad\hat{\vec P}\;.
\label{kinem}
\end{equation}
The remaining generators are Hamiltonians (interaction
depending):
\begin{equation}
\hat P^0\>,\quad \hat{\vec N}\;.
\label{hamil}
\end{equation}

The additive including of interaction into the mass
square operator (Bakamjian--Thomas procedure \cite{BaT53},
see, e.g.,~\cite{KeP91} for details)
presents one of the possible technical ways to  include
interaction in the algebra
(\ref{algebra}):
\begin{equation}
\hat M_0^2 \to \hat M_I^2 = \hat M_0^2 + \hat U \;.
\label{M0toMI}
\end{equation}
Here $\hat M_0$ is the operator of invariant mass for
the free system and
$\hat M_I$ -- for the system with interaction. The interaction
operator $\hat U$ has to satisfy the following commutation relations:
\begin{equation}
\left [\hat {\vec P},\,\hat U\right ]
= \left[\hat {\vec J},\,\hat U\right ]
= \left [\vec\bigtriangledown_P,\,\hat U\right ] = 0\;.
\label{[PU]=0}
\end{equation}
These constraints (\ref{[PU]=0}) ensure that the
algebraic relations
(\ref{algebra}) are fulfilled for interacting system.
The constraints (\ref{[PU]=0}) are not too strong. For instance,
a large class of nonrelativistic potential satisfies
(\ref{[PU]=0}).
The relations (\ref{[PU]=0}) mean that the interaction potential
does not depend on the total momentum of the system. This fact
is well established for a class of potential, for example,
for separable potentials
\cite{ItB90}. Nevertheless, the conditions
(\ref{M0toMI}) and (\ref{[PU]=0}) can be considered as the model
ones. There exists another approach
\cite{Shi59} where a potential depends on the total momentum
but that approach is out of scope of this paper.

In RHD the wave function of the system of interacting particles
is the eigenfunction of a complete set of commuting operators.
In IF this set is:
\begin{equation}
 {\hat M}_I^2\>,\quad
{\hat J}^2\>,\quad \hat J_3\>,\quad \hat {\vec P}\;.
\label{complete}
\end{equation}
${\hat J}^2$ is the operator of the square of the total
angular momentum. In IF the operators
${\hat J}^2\;,\;\hat J_3\;, \;\hat {\vec P}$
coincide with those for the free system. So, in
(\ref{complete}) only the operator
$\hat M_I^2$ depends on the interaction.

To find the eigenfunctions for the system
(\ref{complete}) one has first to construct the adequate basis
in the state space of composite system. In the case of
two-particle system (for example, quark-antiquark system
$q\,\bar q$) the Hilbert space in RHD is the direct product of
two one-particle Hilbert spaces:
${\cal H}_{q\bar q}\equiv {\cal H}_q\otimes {\cal H}_{\bar q}$.

As a basis in ${\cal H}_{q\bar q}$
one can choose the following set of two-particle state vectors:
$$
|\,\vec p_1\,,m_1;\,\vec p_2\,,m_2\,\!\rangle
= |\,\vec p_1\,m_1\,\!\rangle \otimes|\, \vec p_1\,m_2\,\rangle\;,
$$
\begin{equation}
\langle\,\!\vec p\,,m\,|\,\vec p\,'\,m'\,\!\rangle
= 2p_0\,\delta (\vec p - \vec p\,')\,\delta _{mm'}\;.
\label{p1m1p2m2}
\end{equation}
Here $\vec p_1 \;,\;\vec p_2$ are 3-momenta of particles,
$m_1\;,\;m_2$ --- spin projections on the axis $z$,
$p_0 = \sqrt {\vec p^2 +M^2}\;$,$\;M$ is the constituent mass.

One can choose another basis where the motion of the
two-particle center of mass is separated and where three
operators of the set
(\ref{complete}) are diagonal:
$$
|\,\vec P,\;\sqrt{s},\;J,\;l,\;S,\;m_J\,\rangle ,
$$
$$
\langle\,\vec P,\;\sqrt{s},\;J,\;l,\;S,\;m_J
|\,\vec P\,',\;\sqrt {s'},\;J',\;l',\;S',\;m_{J'}\,\rangle
$$
$$ = N_{CG}\,\delta^{(3)}(\vec P -\vec P\,')\delta(\sqrt{s} - \sqrt{s'})
\delta_{JJ'}\delta_{ll'}\delta_{SS'}\delta_{m_Jm_{J'}}\;,
$$
\begin{equation}
N_{CG} = \frac{(2P_0)^2}{8\,k\,\sqrt{s}}\>,\quad
k = \frac{1}{2}\sqrt{s - 4M^2}\;.
\label{PkJlSm}
\end{equation}
Here $P_\mu = (p_1 +p_2)_\mu$, $P^2_\mu = s$, $\sqrt {s}$
is the invariant mass of the two-particle system,
$l$ --- the orbital angular momentum in the center--of--mass frame (C.M.S.),
$\vec S\,^2=(\vec S_1 + \vec S_2)^2 = S(S+1)\;,\;S$
--- the total spin in C.M.S., $J$ --- the total angular
momentum with the projection $m_J$.

The basis (\ref{PkJlSm}) is connected with the basis
(\ref{p1m1p2m2}) through the Clebsh--Gordan (CG) decomposition
for the Poincar\'e group.
The decomposition of the direct product
(\ref{p1m1p2m2}) of two irreducible representations of the
Poincar\'e group into irreducible representations
(\ref{PkJlSm}) has the following form \cite{KoT72}:
$$
|\,\vec p_1\,,m_1;\,\vec p_2\,,m_2\,\rangle
= \sum |\,\vec P,\;\sqrt {s},\;J,\;l,\;S,\;m_J\,\rangle
$$
$$
\times
\langle Jm_J|S\,l\,m_s\,m_l\,\rangle Y^*_{lm_l}(\vartheta\,,\varphi )
\langle S\,m_S\,|1/2\,1/2\,\tilde m_1\,\tilde m_2\,\rangle
$$
\begin{equation}
\times \langle\,\tilde m_1|\,D^{1/2}(P,p_1)\,|m_1\,\rangle
\langle\,\tilde m_2\,|\,D^{1/2}(P,p_2)\,|m_2\,\rangle .
\label{KGpr}
\end{equation}
Here the sum is over the variables
$\tilde m_1$, $\tilde m_2$, $m_l$, $m_S$, $l$, $S$,
$J$, $m_J$. $\vec p = (\vec p_1 - \vec p_2)/2$, $p = |\vec p|$,
$\vartheta \,,\varphi$ are the spherical angles of the
vector $\vec p$ in the C.M.S.,
$Y_{lm_l}$ - a spherical harmonics (star means the complex conjugation),
$\langle\,S\,m_S\,|1/2\,1/2\,\tilde m_1\,\tilde m_2\,\rangle$
and $\langle Jm_J|S\,l\,m_S\,m_l\,\rangle$ are the
CG coefficients for the group $SU(2)$,
$\langle\,\tilde m|\,D^{1/2}(P,p)\,|m\,\rangle$ -
the three--dimensional spin rotation matrix to be used
for correct relativistic invariant spin addition.

Let us discuss briefly the relativistic properties of spins. It
is known that the Lorentz transformation for spins is momentum
depending (see, e.g., \cite{Nov75}). So, to perform  Lorentz invariant
spin addition for particles with different momenta
$\vec p$ and $\vec p\,'$ one has to "shift" the spins
to the frame where the momenta are equal to one another.
The spin transforms following the so called small group
which is isomorphic to rotation group and thus, the operator of
such a "shift"is a 3-dimensional rotation matrix
$D(\alpha ,\beta ,\gamma)$ . The Euler angles
$\alpha ,\beta ,\gamma $ can be written in terms of the
components of the vectors
$\vec p$ and $\vec p \,'$. In this way the
"transplantation" of spins on one and the same momentum
is realized. To understand what means this "transplantation"
let us consider an example: one particle has the momentum
$\vec p_1$, mass  $M_1$, spin $j$ and spin projection
$m$, while another particle with momentum $\vec p_2$
and mass $M_2$ has no spin. In the case of free particles the
vector state of this system is
\begin{equation}
|\,\vec p_1\,,M_1\,,j\,,m;\,\vec p_2\,,M_2\,\rangle
= |\,\vec p_1\,,M_1\,,j\,,m\rangle \otimes \,|\,\vec p_2\,,M_2\,\rangle .
\label{1j}
\end{equation}
The two--particle vector state in the case when it is the first
particle that has no spin is
\begin{equation}
|\,\vec p_1\,,M_1;\,\vec p_2\,,M_2\,,j\,,m\,\rangle
= |\,\vec p_1\,,M_1\,\rangle\otimes |\,\vec p_2\,,M_2\,,j\,,m\,\rangle .
\label{2j}
\end{equation}
In nonrelativistic angular momentum theory the states
(\ref{1j}) and (\ref{2j}) are identical.
They describe the two--particle system with momenta
$\vec p_1$ and $\vec p_2$ and total spin
$j$ with the projection $m$. In both cases the total spin can be
obtained in simple way and is equal to the spin
$j$ of the first particle for (\ref{1j}) or of the second
particle for (\ref{2j}).

In relativistic theory the states
(\ref{1j}) and (\ref{2j}) differ essentially. The difference
is caused by the fact that the states transform from one
inertial coordinate system to another in different ways.
As was mentioned before the Lorentz transformation for spin
depends on
the particle momentum and the spins in
(\ref{1j}) and (\ref{2j}) correspond to particles with different
momenta. In relativistic case the states
(\ref{1j}) and (\ref{2j}) coincide only if the particles momenta
are equal. In general case, to connect the state vectors
(\ref{1j}) and (\ref{2j}) one has to "shift" the spin, for
example, to shift the spin in
(\ref{2j}) into the frame where the second particle has
the momentum $\vec p_1$. This shifting transformation is realized
by the matrix
$D^j(p_2,p_1)$ which belongs to the small group. Let us consider
the state vector
\begin{equation}
|\,\vec p_1\,,M_1\,\rangle\otimes
D^j(p_2,p_1)\,|\,\vec p_2\,,M_2\,,j\,,m\,\rangle .
\label{2jD}
\end{equation}
Let us remind that a transformation belonging to the small group
does not act on momenta. It is easy to see that the resulting
vector describes the same state as (\ref{1j}) and transforms
from one frame to another in the same way as (\ref{1j}). To show
this let us use the equation $D(p',p'')\,D(p'',p) = D(p',p)$.
Now the following covariant equality is valid:
$$
|\,\vec p_1\,,M_1\,,j\,,m;\,\vec p_2\,,M_2\rangle
$$
\begin{equation}
= \sum _{m'}\,
|\,\vec p_1\,,M_1;\,\vec p_2\,,M_2\,,j\,,m'\rangle\,
\langle m'|D^j(p_2,p_1)\,|m\rangle .
\label{1j=2jD}
\end{equation}
As one can see from (\ref{1j=2jD})
the spin has "changed" the momentum.
As this operation is very important for the understanding
of the parameterization below let us formulate it in other
words, namely, in terms of the generators of the Lorentz
transformations.

The Lorentz--transformation generator for the state
(\ref{1j})  is of the form:
$$
\hat {\vec N} = \hat {\vec N}^j_1 + \hat {\vec N}_2\;,
$$
\begin{equation}
\hat {\vec N}^j = ip_0\,\frac {\partial}{\partial\vec p}
- \frac {[\vec j \vec p]}{p_0 + M}\;,\quad
\hat{\vec N} = ip_0\,\frac {\partial }{\partial \vec p}\;.
\label{1jN}
\end{equation}
If we perform the transformation (\ref{1j=2jD}), the
$D$-matrix transforms the generators in the following way:
\begin{equation}
D^j(p_2,p_1)\,(\hat{\vec N}^j_1 + \hat {\vec N}_2)\,
[D^j(p_2,p_1)]^{-1} = \hat {\vec N}_1 + \hat {\vec N}^j_2\;.
\label{DND}
\end{equation}
It is just the equations (\ref{1j=2jD}) and  (\ref{DND})
(and only these equations) that present the exact statement that
$D$-function shifts spin from one momentum to another one.
Similar equations (however, more complicated) are valid in the
case of two non-zero spins. In the case of spin 1/2
the $D$ -- function \cite{KoT72} has the form
$$
D^{1/2}(p_1,p_2) = \cos\left(\frac{\omega}{2}\right) -
2i(\vec k\vec j)\,\sin\left(\frac{\omega}{2}\right)\;,
\;\vec k = \frac{[\,\vec p_1\,\vec p_2\,]}{|[\,\vec p_1\,\vec p_2\,]|}\;,
$$
\begin{equation}
\omega =
2\arctan \frac {|[\,\vec p_1\,\vec p_2\,]|}{(p_{10} + M_1)(p_{20} +M_2)
- (\vec p_1\vec p_2)}\;.
\label{D1/2}
\end{equation}
We will use it below.

Let us make a remark concerning the invariance of the
decomposition
(\ref{KGpr}).  The total spin
$S$ and the total orbital angular momentum $l$
in (\ref{KGpr}) play the role of invariant parameters of
degeneracy. However, the square of the total spin
$\vec S = (\vec S_1 + \vec S_2)$ is not invariant.
But one can define the total spin square in invariant way
as follows:
$$
\left\{\,[D^{S_1}(p_1,P)]^{-1}\,\vec S_1\,[D^{S_1}(p_1,P)]\right.
$$
\begin{equation}
+ \left. [D^{S_2}(p_2,P)]^{-1}\,\vec S_2\,[D^{S_2}(p_2,P)]\,\right\}^2
= S(S + 1)\;.
\label{Ds1D}
\end{equation}
Here $P$ is the center-of-mass momentum.
One can see that in C.M.S. the definition (\ref{Ds1D})
coincides with the definition
(\ref{PkJlSm}).  Similarly, one can define the orbital moment
$l$ in invariant way.

The described spin rotation effect in
(\ref{KGpr}) is a purely relativistic effect. If one takes it
into account, one obtains interesting observable
effects \cite{KrT99}.

To obtain the basis vectors
(\ref{PkJlSm}) in terms of vectors (\ref{p1m1p2m2})
one has to inverse (\ref{KGpr}).  The final equation has the
form:
$$
|\,\vec P,\;\sqrt {s},\;J,\;l,\;S,\;m_J\,\rangle
$$
$$
= \sum _{m_1\>m_2}\,\int \,\frac {d\vec p_1}{2p_{10}}\,
\frac {d\vec p_2}{2p_{20}}\,|\,\vec p_1\,,m_1;\,\vec p_2\,,m_2\,\rangle
$$
\begin{equation}
\times
\langle\,\vec p_1\,,m_1;\,\vec p_2\,,m_2\,|
\,\vec P,\;\sqrt{s},\;J,\;l,\;S,\;m_J\,\rangle\;.
\label{Klebsh}
\end{equation}
Here
$$
\langle\,\vec p_1\,,m_1;\,\vec p_2\,,m_2\,|
\,\vec P,\;\sqrt {s},\;J,\;l,\;S,\;m_J\,\rangle
$$
$$
=\sqrt {2s}[\lambda (s,\,M^2,\,M^2)]^{-1/2}\,
2P_0\,\delta (P - p_1 - p_2)
$$
$$
\times\sum_{\tilde m_1\>\tilde m_2}
\langle\,m_1|\,D^{1/2}(p_1\,P)\,|\tilde m_1\,\rangle
\langle\,m_2|\,D^{1/2}(p_2\,P)\,|\tilde m_2\,\rangle
$$
$$
\times
\sum_{m_l\;m_S}\,\langle1/2\,1/2\,\tilde m_1\,\tilde m_2\,|S\,m_S\,\rangle
Y_{lm_l}(\vartheta \,,\varphi)
$$
$$
\times
\langle S\,l\,m_s\,m_l\,|Jm_J\rangle\;.
$$
Here $\lambda (a,b,c) = a^2 + b^2 + c^2 - 2(ab + bc + ac)$.
To obtain (\ref{Klebsh}) the decomposition in terms of spherical
harmonics and the summation of all of the momenta to give the
total momentum $J$ were performed in the C.M.S.
and then the obtained result was shifted to arbitrary coordinate
frame by use of $D$-functions.

It is on the vectors (\ref{PkJlSm}), (\ref{Klebsh})
that the Poincar\'e--group representation is realized
in the vector state space of two free particles.
The vector in representation is determined by
the eigenvalues of the complete commuting set of operators:
\begin{equation}
\hat M^2_0 = \hat P^2\,,\;\hat J^2\,,\;\hat J_3\;.
\label{M0,J}
\end{equation}
The parameters $S$ and $l$
(as was mentioned) play the role of invariant parameters of degeneracy.

As in the basis
(\ref{PkJlSm}) the operators ${\hat J}^2\;,\;\hat J_3\;,\;\hat {\vec P}$
in (\ref{complete}) are diagonal, one needs to diagonalize only the operator
$\hat M_I^2$ in order to obtain the system wave function.

The eigenvalue problem for the operator ${\hat M}^2_I$ in the basis
(\ref{PkJlSm}) has the form of nonrelativistic Schr\"odinger
equation (see, e.g., \cite{KeP91}).

The corresponding composite--particle wave function has the form
$$
\langle\vec P\,',\,\sqrt {s'},\,J',\, l',\,S',\,m_J'|\,p_c\rangle  =
$$
\begin{equation}
=N_C\,\delta (\vec P\,' - \vec p_c)\delta _{JJ'}\delta _{m_Jm_J'}\,
\varphi^{J'}_{l'S'}(k')\;,
\label{wf}
\end{equation}
$$
N_C =
\sqrt{2p_{c0}}\sqrt{\frac{N_{CG}}{4\,k'}}\;,
$$
$|\,p_c\rangle$ is an eigenvector of the set (\ref{complete});
$J(J+1)$ and $m_J$ are the eigenvalues of $\hat J^2$, $\hat J_3$,
respectively (Eqs.~(\ref{complete}), (\ref{M0,J})).

The two--particle wave function of relative motion
for equal masses and total angular momentum and total
spin fixed is:
\begin{equation}
\varphi^J_{lS}(k(s)) = \sqrt[4]{s}\,u_l(k)\,k\;,
\label{phi(s)}
\end{equation}
and the normalization condition has the form:
\begin{equation}
\sum_l\int\,u_l^2(k)\,k^2\,dk = 1\;.
\label{norm}
\end{equation}

Let us note that for composite quark systems one uses sometimes
instead of equation (\ref{norm})
the following one:
\begin{equation}
 n_c\,\sum_l\int\,u_l^2(k)\,k^2\,dk = 1\;.
\label{norm nc}
\end{equation}
Here $n_c$ -- is the number of colours.
The wave function (\ref{phi(s)})
coincides with that obtained by "minimal relativization" in
\cite{FrG89}.
The normalization factors in
(\ref{phi(s)})
in this case correspond to the relativization obtained by the
transformation to relativistic density of states
\begin{equation}
k^2\,dk\quad\to\quad \frac{k^2\,dk}{2\sqrt{(k^2 + M^2)}}\;.
\label{rel den}
\end{equation}

It is worth to notice that wave functions in RHD (for example,
the wave function
(\ref{wf}) defined as the eigenfunction of the operators set
(\ref{complete})) in general are not the same as relativistic
covariant wave functions defined as solutions of wave equations
or as the matrix elements of local Heisenberg field.

The formalism of this Section is used in the next one to
present the method of calculation of electroweak properties of
composite systems. Particularly, the method of construction of
electroweak current operators is described.

\section{The new relativistic instant--form approach to the
electroweak structure of two body composite systems.}

In this Section we present our approach to electroweak
properties of relativistic two--particle systems. To demonstrate
how one describes the electromagnetic properties of composite
systems in our version of the RHD instant form we first use the
following simple model. We consider the system of two
spinless particles in the
$S$-- state of relative motion, one particle having no charge.
Let us note that a similar model was used in
\cite{KeP91} where the authors gave the description of
constituent interaction in IF of RHD and obtained the mass
spectrum. The application of our method in general case
follows the scheme of this Section.
The case of $\pi$ - meson is investigated in Sec.IV and the $S=1$
case in \cite{KrT01}.

Electromagnetic properties of the system are determined by the
current operator matrix element. This matrix element is
connected with the charge form factor
$F_c(Q^2)$ as follows:
\begin{equation}
\langle p_c\,|j_\mu(0)|\,p'_c\,\rangle  = (p_c+p'_c)_\mu\,F_c(Q^2)\;,
\label{j=Fc}
\end{equation}
where $p'_c\;,\;p_c$ are 4--momenta of the composite system in
initial and final states,
$Q^2 = -t\;,\;q^2 = (p_c - p_c')^2 = t\;,\;q^2$ is the momentum--transfer
square.
The form ~(\ref{j=Fc}) is defined by the Lorentz covariance and
by the conservation law only and does not depend on the model
for the internal structure of the system.

The Eq. ~(\ref{j=Fc}) presents the simplest example of the
extraction of a reduced matrix element. The 4--vector
$(p_c + p'_c)_{\mu}$ describes symmetry and
transformation properties of the matrix element. The reduced
matrix element (the form factor) contains all the dynamical
information on the process described by the current. Usually,
one does not fix the dependence of form factor on a scalar
mass of the composite system
${p_c}^2 = {p'_c}^2 = {M_c}^2$,
because it is diagonal with respect to this variable.
The representation of a matrix element in terms of form factors
often is referred to as the parameterization of matrix element.
The scattering cross section for elastic scattering of electrons
by a composite system can be expressed in terms of charge form
factor
$F_c(Q^2)$. So, form factor can be obtained
from experiment and it is interesting to calculate it in a
theoretical approach.

In this Section we calculate the form factor of our simple
composite system using the version of RHD IF based on the
approach of the Section II.

Now let us list the conditions for the operator of the conserved
electromagnetic current to be fulfilled in relativistic case
(see, e.g.,~\cite{Lev95}).\\
(i).{\it Lorentz--covariance}:
\begin{equation}
\hat U^{-1}(\Lambda )\hat j^\mu (x)\hat U(\Lambda ) =
\Lambda ^\mu_{\>\nu}\hat j^\nu (\Lambda ^{-1}x)\;.
\label{UjmuU}
\end{equation}
Here $\Lambda $ is the Lorentz--transformation matrix,
$\hat U(\Lambda ) $ -- the operator of the unitary
representation of the Lorentz group.\\
(ii).{\it Invariance under translation}:
\begin{equation}
\hat U^{-1}(a)\hat j^\mu(x)\hat U(a) = \hat j^\mu(x-a)\;.
\label{UajmuUa}
\end{equation}
Here $\hat U(a)$  is the operator of the unitary
representation of the translation group.\\
(iii).{\it Current conservation law}:
\begin{equation}
[\,\hat P_\nu\,\hat j^\nu(0)\,] = 0\;.
\label{Pj=0}
\end{equation}
In terms of matrix elements
$\langle\,\hat j^\mu(0)\,\rangle$
the conservation law can be written in the form:
\begin{equation}
q_\mu\,\langle\,\hat j^\mu(0)\,\rangle  = 0\;.
\label{Qj=0}
\end{equation}
Here $q_\mu$ is 4-vector of the momentum transfer.\\
(iv).{\it Current--operator transformations under space--time
reflections are}:
$$
\hat U_P\left(\,\hat j^0(x^0\,,\vec x)\,,\hat{\vec j}(x^0\,,\vec x)\right)
\hat U^{-1}_P
$$
$$
=
\left(\,\hat j^0(x^0\,,-\,\vec x)\,,-\,\hat{\vec j}(x^0\,,-\,\vec x)\right)\;,
$$
\begin{equation}
\hat U_R\,\hat j^\mu(x)\,\hat U^{-1}_R = \hat j^\mu(-\,x)\;.
\label{UpUr}
\end{equation}
In (\ref{UpUr}) $\hat U_P$ is the unitary operator for the
representation of space reflections and
$\hat U_R$ is the antiunitary operator of the representation of
space-time reflections
$R = P\,T$.\\
(v).{\it Cluster separability condition}:
If the interaction is switched off then the current operator
becomes equal to the sum of the operators of one--particle
currents.\\
(vi).{\it The charge is not renormalized by the interaction
including}:  The electric charge of the system with interaction
is equal to the sum of the constituent electric charges.

In this paper the explicit equations for the form factors are
obtained taking into account all the listed conditions.

\subsection* {A. Electromagnetic properties of the system of
free particles}

Let us consider first the simple two--particle system 
described in the beginning of Section III.
The elastic scattering of a test particle, e.g., of an
electron, by the system are defined by the operator of
electromagnetic current $j^{(0)}_\mu(0)$ of the two--particle
free system. This operator can be calculated in the
representation given by the basis
(\ref{p1m1p2m2}) or in the representation given by the basis
(\ref{PkJlSm}).
In the first case the operator has the form
$j^{(0)}_\mu = j_{1\mu}\otimes I_{2}$. Here
$j_{1\mu}$ is the electromagnetic current of the charged particle and
$I_{2}$ is the unity operator in the Hilbert space of
states of the uncharged particle.
$$
\langle\vec p_1;\vec p_2|j_\mu^{(0)}(0)|\vec p\,'_1;\vec p\,'_2\rangle
$$
\begin{equation}
= \langle\vec p_2|\vec p\,'_2\rangle
\langle\vec p_1|j_{1\mu}(0)|\vec p\,'_1\rangle\;.
\label{j=j1xI}
\end{equation}
The matrix element of the one spinless particle current
in the free case contains only one form factor -- the charge form
factor of the charged particle
$f_1(Q^2)$:
\begin{equation}
\langle\vec p_1|j_{1\mu}(0)|\vec p\,'_1\rangle  =
(p_1+p'_1)_\mu\,f_1(Q^2)\;.
\label{j=f1}
\end{equation}

So, the electromagnetic properties of the system of two free
particles
(\ref{j=Fc}) are defined by the form factor
$f_1(Q^2)$, containing all the dynamical information on elastic
processes described by the matrix element
(\ref{j=j1xI}) \cite{KeP91}.
Particularly, the charge of the system is defined by the value
of this form factor at $Q^2\to 0$:
\begin{equation}
\lim_{Q^2\to 0}\,f_1(Q^2) = f_1(0) = e_c\;.
\label{f1(0)}
\end{equation}
$e_c$ is the system charge.

Now let us write the electromagnetic--current matrix element
for the two--particle free system in the basis where the
center--of--mass motion is separated (\ref{PkJlSm}):
\begin{equation}
\langle\vec P,\sqrt s,\mid j_\mu^{(0)}(0) \mid \vec P',\sqrt{s'}\rangle\;.
\label{*}
\end{equation}
Here the variables which take zero values are omitted:
$J = S = l = 0$. One can consider the matrix element
(\ref{*}) as a matrix element of an irreducible tensor operator
on the Lorentz group and one can apply the Wigner--Eckart
theorem. Under the condition of the theorem the matrix element
of an irreducible tensor operator is the product of two factors:
the invariant part (reduced matrix element) and the CG
coefficient which defines the transformation properties of the
matrix element. Thus, one can write
(\ref{*}) in the form
$$
\langle\vec P,\sqrt s,\mid j_\mu^{(0)}(0) \mid \vec P',\sqrt{s'}\rangle =
$$
\begin{equation}
=  A_\mu (s,Q^2,s')\;g_0 (s,Q^2,s')\;.
\label{j=A mu g0}
\end{equation}
The motivation for the parameterization
(\ref{j=A mu g0}) is easy to be understood for our simple
system. The 4--vector $A_{\mu}$ describes the transformation
properties of the matrix element and the invariant function
$g_0 (s,Q^2,s')$ contains the dynamical information on the
process. We will refer to $g_0 (s,Q^2,s')$ as to free
two--particle form factor. For more complicated systems the
parameterization corresponding to the Wigner--Eckart theorem
for the Lorentz group can be performed using a special
mathematical techniques as described in the papers
~\cite{ChS63},~\cite{KoT72}.

The vector $A_\mu (s,Q^2,s')$ which describes the matrix--element
transformation properties is defined by the 4--momenta of
initial and final states only: we have no other vectors to our
disposal. So $A_{\mu} (s,Q^2,s')$ is a linear combination of
4--momenta of initial and final states and is defined by the
current transformation properties (the Lorentz--covariance and
the conservation law):
\begin{equation}
A_\mu =\frac{1}{Q^2}[(s-s'+Q^2)P_\mu + (s'-s+Q^2) P\,'_\mu]\;.
\label{Amu}
\end{equation}

Thus, in the basis (\ref{PkJlSm}) the electromagnetic properties
of the free two--particle system are defined by the free two--
particle form factor
$g_0 (s,Q^2,s')$.

So, in both representations (defined by the basis
(\ref{p1m1p2m2}) as well as by the basis (\ref{PkJlSm}))
we pass from the description of the system in terms of matrix
elements to that in terms of Lorentz--invariant form factors.

One can see that (\ref{j=j1xI}) and
(\ref{j=A mu g0}) describe electromagnetic properties in terms
of only one form factor. Both of these descriptions are,
certainly, equivalent from the physical point of view.
Let us consider the difference between these descriptions.
As we will show below by direct calculation the free
two--particle form factor
$g_0 (s,Q^2,s')$ is not an ordinary function but has to be
considered in the sense of distributions
in variables $s\;,\;s'$, generated by a locally integrable
function. So, $g_0 (s,Q^2,s')$ is a regular
generalized function. Let us remind that regular generalized
function is that defined through an integral in the space of
test functions. So, all the properties of
$g_0(s,Q^2,s')$  have to be considered as the properties of a
functional given by the integral over the variables
$s\;,\;s'$ of the function $g_0(s,Q^2,s')$
multiplied by a test function. As test functions it is
sufficient to take a large class of smooth functions that give
the uniconvergence of the integral. In particular, the limit
(\ref{f1(0)}) giving the total charge of the system through
two--particle form factor is now the weak limit:
\begin{equation}
\lim_{Q^2\to 0}\langle g_0(s,Q^2,s')\,,\phi(s,s')\rangle .
\label{g0(0)}
\end{equation}
Here $\phi(s,s')$ is a function from the space of test
functions. The precise definition of the functional will
be given below.

As the invariant variables $s\>,\>s'$ contain the energies of
the relative motion of particles in initial and final states,
one can consider the integral in
(\ref{g0(0)}) as an integral over these energies.

At the first glance it seems that the description of the
two--particle free system in terms of the form factor
$g_0(s,Q^2,s')$ is too complicated. However, so is the reality,
as we will see later in the Subsection III.F. In fact, this kind
of description is used implicitly for a long time in
nonrelativistic theory of composite systems,
without calling things by their proper names. It is this kind
of description that makes it possible to construct the
electromagnetic current operator with correct transformation
properties for interacting systems.


\subsection* {B. The form factor of the system of two free
particle.}

The locally integrable function
$g_0(s,Q^2,s')$ can be easily obtained by use of CG
decomposition (\ref{Klebsh}) for the Poincar\'e group.
Using (\ref{Klebsh})  we obtain for (\ref{j=A mu g0}):
$$
\langle\vec P,\sqrt{s},\mid j_\mu^{(0)}(0) \mid \vec P',\sqrt{s'}\rangle
$$
$$
=  \int\,\frac{d\vec p_1}{2\,p_{10}}
\frac{d\vec p_2}{2\,p_{20}}\frac{d\vec p_1\,'}{2\,p'_{10}}
\frac{d\vec p_2\,'}{2\,p'_{20}}\,
\langle\vec P,\sqrt s,\mid \vec p_1;\vec p_2\rangle
$$
\begin{equation}
\times\langle\vec p_1;\vec p_2|j_\mu^{(0)}(0)|\vec p\,'_1;\vec p\,'_2\rangle
\langle\,\vec p\,'_1;\vec p\,'_2|\vec P',\sqrt{s'}\rangle\;.
\label{jP=jp1p2}
\end{equation}
To calculate the free two--particle form factor one has to use
(\ref{j=j1xI}), (\ref{j=f1}), (\ref{j=A mu g0})
and the explicit form of CG coefficients
(\ref{Klebsh}) for quantum numbers of the system.
As the particles of the system under consideration are spinless,
now (\ref{Klebsh}) does not contain $D$ -- functions.

It is convenient to integrate in (\ref{jP=jp1p2})
using the coordinate frame with
$\vec P\,'=0\;,\;\vec P =(0,0,P)$.
As the result we obtain the following relativistic invariant
form for the function
$g_0(s,Q^2,s')$:
$$ g_0(s,Q^2,s')
$$
\begin{equation}
= \frac{(s+s'+Q^2)^2\,Q^2}{2\,\sqrt{(s-4M^2) (s'-4M^2)}}\>
\frac{\vartheta(s,Q^2,s')}{{[\lambda(s,-Q^2,s')]}^{3/2}}\,f_1(Q^2)\;.
\label{ff-nonint}
\end{equation}
Here $\vartheta(s,Q^2,s')=
\theta(s'-s_1)-\theta(s'-s_2)$, and $\theta$ is the step
function.
The result, naturally, does not depend on the choice of the
coordinate frame.

$$
s_{1,2}=2M^2+\frac{1}{2M^2} (2M^2+Q^2)(s-2M^2)
$$
$$
\mp \frac{1}{2M^2}\sqrt{Q^2(Q^2+4M^2)s(s-4M^2)}\;.
$$
The functions $s_{1,2}(s,Q^2)$ give in the plane
$(s,s')$ the kinematically available region.
The position of this region depends strongly on the
momentum--transfer square
$t = -Q^2$
\footnote{
The \protect$\vartheta$ -- function is a purely kinematical factor for
IA. This fact does not depend on relativism and takes place in
nonrelativistic case, too. See Subsection III.F for details.}.
The simplest way to obtain the functions $s_{1,2}$
is a geometrical one \cite{TrS69}.
\begin{figure*}
\centerline{\epsfxsize=0.4\textwidth \epsfbox{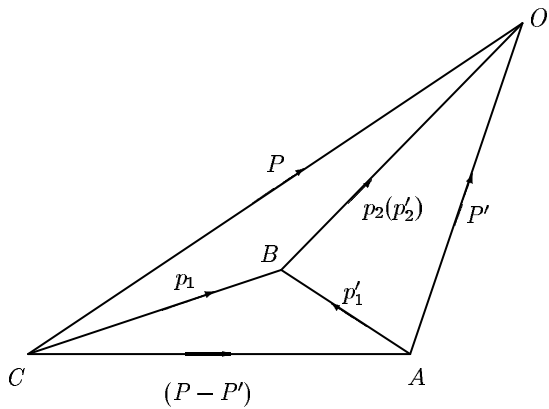}}
\caption{Kinematical triangle}
\end{figure*}

Let us construct the triangle schematically presented on the
Fig.1. The side $OB$ presents the vector
$p_{2\,\mu}$, and $p'_{2\,\mu}$, as following (\ref{j=j1xI})
$p_{2\,\mu} = p'_{2\,\mu}$. The side $CB$ presents the vector
$p_{1\,\mu}$ and $AB$ -- $p'_{1\,\mu}$. Now the sides of the
large triangle $AOC$ present the vectors
$P_\mu'\>,\>P_\mu$ and $(P_\mu - P'_\mu)$, with the norms
$\sqrt{s'},\sqrt{s}, \sqrt{t}$, respectively.
Vectors of initial state $p'_{1\,\mu}\;$,$\;p_{2\,\mu} = p'_{2\,\mu}$,
$\;P_\mu'$ are fixed. Let us fix the norm of vector $(P_\mu - P'_\mu)$.
So , because the norm of vector $p_{1\,\mu}\;,\;p_{1\,\mu}^2= M^2$ is
constant,
the triangles $ABO$ and $ABC$ are determined unambiguously by three
sides. But the triangle $ABC$ can be rotated around the side
$AB$ ($p'_{1\,\mu}$).  It is possible to find the minimal
$\sqrt{s_1}$ and maximal $\sqrt{s_2}$ lengths of $OC$ (norm of
vector $P_\mu$) under this rotation. The value of the
$s_1\;,\;s_2$ give the kinematically available region in the
plane $(s\;,\;s')$ which is symmetric under interchange
$s\leftrightarrow s'$.

\begin{figure*}
\centerline{\epsfxsize=0.4\textwidth \epsfbox{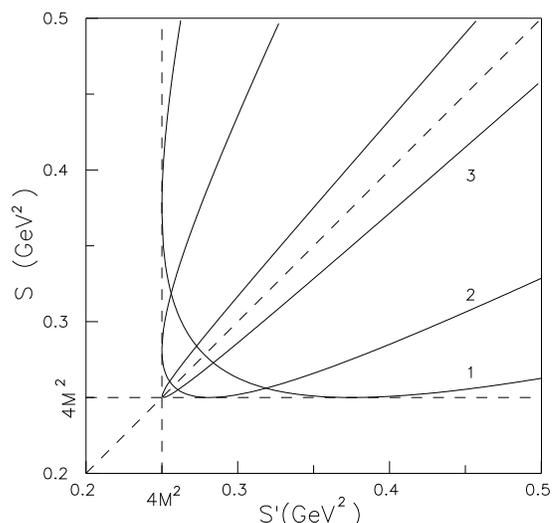}}
\caption{The kinematically available
region
in the plane $s, s'$ (inside the parabolae). The calculation
is performed for: 1) $Q^2 = 2 M^2$. 2) \protect$Q^2 = M^2/2$. 3)
\protect$Q^2 = M^2/64$. The constituent mass is
$M$ = 0.25 GeV.}
\end{figure*}

In Fig.2 the domain where the generalized function
(\ref{ff-nonint}) is nonzero in the plane $s\>,\>s'$
is given for different values of the momentum--transfer square.
One can see that the free two--particle form factor
$g_0(s,Q^2,s')$ (\ref{ff-nonint}) has in fact to be interpreted
in terms of the distributions: The ordinary limit as
$Q^2\to$ 0 is zero because of the cutting
$\vartheta$ -- functions and the static limit exists only as
the weak limit (\ref{g0(0)}).

Let us calculate this limit. Let us define the functional giving
regular generalized function as a functional in
${\bf R}^2$ as follows:
\begin{equation}
\langle\tilde g_0 (s,Q^2,s')\,,\phi(s,s')\rangle  = \int\,d\mu(s,s')\,
\tilde g_0 (s,Q^2,s')\,\phi(s,s')\;.
\label{<>}
\end{equation}
Here
$$
\tilde g_0 (s,Q^2,s')
$$
\begin{equation}
= 16\,\theta(s - 4M^2)\,
\theta(s' - 4M^2)\,g_0 (s,Q^2,s')\;.
\label{tilde g0}
\end{equation}
\begin{equation}
d\mu(s,s') =
\sqrt[4]{ss'}\,d\mu(s)\,d\mu(s')\>,\quad
d\mu(s) = \frac{1}{4}k\,d\sqrt{s}\;.
\label{dmu}
\end{equation}
The $\theta$ -- functions in
(\ref{tilde g0}) give the physical region of possible variations
of the invariant mass squares in the initial and final states
explicitly.
The measure
(\ref{dmu}) is due to the relativistic density of states
(\ref{phi(s)}), (\ref{rel den}).  $\phi(s\,,s')$
is a function from the test function space. So, for example, the
limit of $g_0(s\,,Q^2\,,s')$ as $Q^2\;\to\;$0  (the static
limit)  has the meaning only as the weak limit
(compare with (\ref{f1(0)})):
\begin{equation}
\lim_{Q^2\to 0}\langle\,\tilde g_0,\phi\,\rangle
= \langle{e}_á \delta(\mu(s') - \mu(s))\vartheta(s - 4\,M^2), \phi\,\rangle\;.
\label{Q2=0}
\end{equation}
It is this weak limit that gives the electric charge of the free
two--particle system. If the test functions are normalized with
the relativistic density of states, then the r.h.s. of the Eq.
(\ref{Q2=0}) is equal to the total charge of the system.


\subsection*{C.
Electromagnetic structure of the system of two interacting
particles.}

Now let us consider the electromagnetic structure of our simple
model (\ref{j=Fc}) in the case of interacting particles.

As we have mentioned in Sec.II when constructing the bases
(\ref{p1m1p2m2}) and (\ref{PkJlSm}) in the frame of RHD the
state vector
$|\,p_c\,\rangle $ belongs to the direct product of two
one--particle spaces. We can write the decomposition of this
vector with $J=l=S=m_J=$0 in the basis
(\ref{PkJlSm}). Now (\ref{j=Fc}) has the form:
$$
\int\,\frac{d\vec P\,d\vec
P\,'}{N_{CG}\,N_{CG}'}\,d\sqrt{s}\,d\sqrt{s'}\,
\langle p_c|\vec P\,,\sqrt{s}\,\rangle
\langle\vec P\,,\sqrt{s}|j_\mu(0)|\vec P\,'\,,\sqrt{s'}\rangle
$$
\begin{equation}
\times
\langle\vec P\,'\,,\sqrt{s'}|p_c'\rangle  = (p_c+p'_c)_\mu\,F_c(Q^2)\;.
\label{int=Fá}
\end{equation}
Here
$\langle\vec P\,'\,,\sqrt{s'}|p_c'\rangle$
is the wave function in the sense of the instant form of RHD
(\ref{wf}).

Using (\ref{wf}) we obtain for (\ref{int=Fá}):
$$
\int\,\frac{N_c\,N'_c}{N_{CG}\,N_{CG}'}\,d\sqrt{s}\,d\sqrt{s'}\,
\varphi(s)\,\varphi(s')
$$
\begin{equation}
\times \langle\vec p_c\,,\sqrt{s}|j_\mu(0)|\vec p_c\,'\,,\sqrt{s'}\rangle
= (p_c+p'_c)_\mu\,F_c(Q^2)\;.
\label{int ds=Fc}
\end{equation}
We have omitted in the wave function
(\ref{phi(s)}) the variables with zero values:
$J=S=l=$0 (see (\ref{*}) too).

Using (\ref{phi(s)}), (\ref{dmu}) we can rewrite (\ref{int ds=Fc})
in the form of the functional in ${\bf R}^2$:
$$
\int\,d\mu(s\,,\,s')\,
u(k(s))\,J_\mu(\vec p_c\,,\sqrt{s};\vec p_c\,'\,,\sqrt{s'})\,u(k(s'))
$$
\begin{equation}
= (p_c+p'_c)_\mu\,F_c(Q^2)\;,
\label{int dmu=Fá}
\end{equation}
$$
J_\mu(\vec p_c\,,\sqrt{s};\vec p_c\,'\,,\sqrt{s'})
= 16\,\theta(s - 4\,M^2)\,\theta(s' - 4\,M^2)\,
$$
$$
\times
\frac{N_c\,N'_c}{N_{CG}\,N_{CG}'}
\langle\vec p_c\,,\sqrt{s}|j_\mu|\vec p_c\,'\,,\sqrt{s'}\rangle\;.
$$

In the previous cases the state vectors
and the operators entering matrix elements transformed following
one and the same representation of the Poincar\'e group (namely,
following the universal covering subgroup of the Poincar\'e
group -- nonuniform group $SL(2,C)$
\cite{Nov75}). Now in the matrix element in the integrand of
(\ref{int dmu=Fá}) the state vectors and the operator transform
following the different representations of this group. The
current operator describes the transitions in the system of two
interacting particles and transforms following the
representation with the generators of Lorentz boosts depending
on the interaction (\ref{complete}). The state vectors belong to
the basis (\ref{PkJlSm}) and physically describe the system of
two free particles and, so, transform following a representation
with generators which do not depend on the interaction
(\ref{M0,J}). That is why the current operator matrix element
in (\ref{int dmu=Fá}) can not be represented in the form
(\ref{j=A mu g0}), (\ref{Amu}): we can not construct the
4--vector defining the matrix--element transformation properties
under Lorentz boosts from the variables which the state vectors
depend on.

Nevertheless, as we show below, the problem of the
parameterization of the current matrix element in
(\ref{int dmu=Fá}) can be solved if one consider this equality
as the equality of two functionals. The l.h.s. contains a
functional in
$R^2$ generated by the Lorentz--covariant function (current
matrix element). Let us denote
\begin{equation}
\psi(s\,,\,s') = u(k(s))\,u(k'(s'))\;.
\label{psi(ss')}
\end{equation}
The functional in the l.h.s. of (\ref{int dmu=Fá})
is given on the set of test functions
$\psi(s\,,\,s')$ through an integral in $R^2$
and defines a Lorentz--covariant (regular) generalized function
with the values in the Minkowski space
(see, e.g., \cite{BoL87}). Here $Q^2$ is a parameter.
The test--function space can be (in general) larger than
(\ref{psi(ss')}). However, the uniconvergence of
(\ref{int dmu=Fá}) has to be guaranteed.

Let us write the matrix element in the form analogous to
(\ref{j=A mu g0}):
\begin{equation}
J_\mu(\vec p_c\,,\sqrt{s};\vec p_c\,'\,,\sqrt{s'})
= B_\mu(s\,,Q^2\,,s')G(s\,,Q^2\,,s')\;.
\label{j=BG}
\end{equation}
The covariant part in (\ref{j=BG}) (as well as in
(\ref{j=A mu g0})), the vector $B_\mu(s\,,Q^2\,,s')$,
is supposed to be an ordinary smooth function and the invariant
part
$G(s\,,Q^2\,,s')$ is generalized function. In fact,
$G(s\,,Q^2\,,s')$ is the reduced matrix element containing the
information on the process. This kind of representation of a
Lorentz covariant generalized function as a product of a Lorentz
covariant ordinary smooth function and a Lorentz invariant
generalized function was described in
\cite{BoL87}.  Let us remark that all normalization
constants enter the vector $B_\mu$. Contrary to (\ref{j=A mu
g0}), now it is impossible to construct the 4--vector
$B_\mu$, describing the transformation properties of the matrix
element, in terms of the variables entering the state vectors
(as was pointed out before).

Using (\ref{j=BG}) we can rewrite (\ref{int dmu=Fá}) in the
following form:
$$
\int\,d\mu(s\,,\,s')\,\psi(s\,,s')B_\mu(s\,,Q^2\,,s')G(s\,,Q^2\,,s')
$$
\begin{equation}
= (p_c + p'_c)_\mu\,F_c[\psi](Q^2)\;.
\label{int BG=Fá}
\end{equation}
To obtain the vector $B_\mu$ let us require the
Eq.(\ref{int BG=Fá}) to be covariant in the sense of
distributions, that is to be valid for any test function
$\psi(s\,,s')$ in any fixed frame.
The variation of test function in the functional
(\ref{int BG=Fá}) means in fact, following
(\ref{psi(ss')}), the variation of the wave function of the
internal motion. Under such a variation the vector in the r.h.s
of (\ref{int BG=Fá}) is unchanged as it is constructed with
4--vectors describing the motion of the system as a whole,
independent of the internal constituent motion. As to the form
factor in the r.h.s it varies under the test function variation.
So, under a variation of the test function the r.h.s. of
(\ref{int BG=Fá}) remains to be collinear to the vector
$(p_c + p'_c)_\mu$. At the same time, under arbitrary variation
of the test function the vector in the l.h.s. in general changes
the direction. So, for the validity of the equality
(\ref{int BG=Fá}) with arbitrary test function it is sufficient
to require that the following equation
\begin{equation}
B_\mu(s\,,Q^2\,,s') = (p_c + p'_c)_\mu
\label{B}
\end{equation}
holds. This choice of the vector $B_\mu$ in (\ref{B})
ensures that the l.h.s. of
(\ref{int dmu=Fá}) satisfies the condition of Lorentz covariance
for the current as well as the condition of current
conservation.

Let us discuss the physical meaning of the representation
(\ref{j=BG}), (\ref{B}) for the matrix element. As this
representation is explicitly Lorentz covariant and also
satisfies the current conservation law, then it means that the
current operator for the composite system contains the
contribution not only of one--particle currents but of
two--particle currents, too
(see, e.g., \cite{KeP91}):
\begin{equation}
j(x) = \sum_{k}\,j^{(k)}(x) + \sum_{k<m}\,j^{(km)}\;.
\label{j=jk+jkm}
\end{equation}
Here the first term is the sum of one--particle currents and the
second -- of two--particle currents. In the case of our simple
model each sum in (\ref{j=jk+jkm})
contains only one term. It is well known that if one approximates
$j(x) \approx \sum_{k}\,j^{(k)}(x)$  then the current operator
in IF dynamics does not satisfy the condition of Lorentz
covariance and the conservation law
~\cite{KeP91}. So, from the physical point of view, the
covariant part of the current matrix element
(\ref{B}) which defines the transformation properties of the
current in (\ref{int dmu=Fá}) is given by
(\ref{j=jk+jkm}) and contains the contributions of one-- and
two-- particle currents.

The invariant part of the decomposition
(\ref{j=BG}) is the form factor or the reduced matrix element
$G(s\,,Q^2\,,s')$ and contains the information on the dynamics of
the scattering of test particle by each of the constituents
(the first term in (\ref{j=jk+jkm})), i.e. by the free
two--particle system, as well as by two constituent
simultaneously (the second term). So, the form factor contains
the contribution of the free system form factor
(\ref{ff-nonint}) and the contribution of some exchange currents
analogous to meson currents in nucleon systems \cite{GrR87}.

\begin{equation}
G(s\,,Q^2\,,s') = g_0(s,Q^2,s') + G_c(s\,,Q^2\,,s')\;.
\label{G=g0+Gc}
\end{equation}
Here $G_c$ is the reduced matrix element containing the
contribution of two--particle currents
~(\ref{j=jk+jkm}).

Using
(\ref{phi(s)}), (\ref{dmu}), (\ref{psi(ss')}), (\ref{B})
one can obtain from (\ref{int BG=Fá}) the scalar equation
of the following form:
\begin{equation}
\int\,d\sqrt{s}\,d\sqrt{s'}\,
\varphi(s)\,G(s\,,Q^2\,,s')\, \varphi(s') = F_c(Q^2)\;.
\label{int G=Fá}
\end{equation}
The form factor $G(s\,,Q^2\,,s')$ includes all possible
mechanisms of the transition described by the matrix element
(\ref{j=Fc}). So, the representation
(\ref{int G=Fá}) for the charge form factor of the system is
quite general.

Now let us proceed with the approximate calculation of the form factor
(\ref{int G=Fá}).

\subsection*{D.
Modified impulse approximation
(MIA)}

The problem of the calculation of the form factor
$G(s\,,Q^2\,,s')$ (\ref{int G=Fá}) including exchange currents
is a very difficult problem. We propose an approximation which
is a kind of analog of relativistic impulse approximation. We
propose to omit the contribution of the two--particle currents
to the form factor $G(s\,,Q^2\,,s')$.

However we will not change the covariant part
$B_{\mu}$ of the current matrix element in
(\ref{j=BG}), so that this covariant part will contain the
contribution of the two--particle currents and so that the
transformation properties of the matrix element will not be
changed.

So, we approximately change the generalized function
$G(s\,,Q^2\,,s')$
in (\ref{j=BG}), (\ref{G=g0+Gc})
for the generalized function $g_0 (s,Q^2,s')$
(\ref{j=A mu g0}), (\ref{ff-nonint}),
which describes, as we have shown before, the electromagnetic
properties of the free two--particle system. Nevertheless,
the matrix element
(\ref{int=Fá}), (\ref{j=BG})
as a whole will contain the contributions of two--particle
currents, although not the full contribution but such that
ensures its correct transformation properties.

Let us note that our approximation does not contradict general
statements (see ~\cite{KeP91}) that to obtain correct
description of electromagnetic current of composite system which
satisfy the Lorentz--covariance condition and the current
conservation law one has to take into account many--particle
currents.

Let us discuss now the meaning of our approximation from the
point of view of the Wigner--Eckart theorem for the Lorentz
group. The matrix element of a current including many--particle
currents, following the Wigner--Eckart theorem for the
Lorentz
group, can be presented in the form
~(\ref{j=BG}), ~(\ref{B}).  The dynamical information on
many--particle currents is contained in the reduced matrix
element -- the form factor, while the transformation properties
of the contributions of many--particle currents are defined by
the covariant part of the form
~(\ref{j=BG}). So, our approximation means that the dynamical
part of the contribution of the many--particle currents to the
total current is omitted while the covariant part of the
contributions remains. The dynamics of many--particle currents
remains out of the limits of the approximation, while the
transformation properties of the total current remain intact.

Thus, in our approximation the scalar equality
(\ref{int G=Fá}) transforms into approximate scalar equality
which corresponds, from the physical point of view, to
relativistic impulse approximation. In the developed
mathematical formalism we have not broke the Lorentz covariance
of the current nor the current conservation law.
Let us point out that to calculate form factor
we do not use a special current component as it is
done in other mathematical formulations of RHD
(see, e.g., \cite{ChC88}). Let us remark that, from the physical
point of view, the form factor $g_0(s,Q^2,s')$ contains the
contributions of one--particle currents only
(see Equations (\ref{j=A mu g0}),
(\ref{jP=jp1p2}), (\ref{ff-nonint})) and in this sense our
approximation corresponds to the known impulse approximation.
In order to emphasize that our approximation differs from the
usual IA we will refer to it as to modified impulse
approximation (MIA). The form factor of the composite system
in MIA has the form:
\begin{equation}
F_á(Q^2) =
\int\,d\sqrt{s}\,d\sqrt{s'}\,\varphi(s)\,g_0(s\,,Q^2\,,s')
\varphi(s')\;.
\label{Fc fin}
\end{equation}

It is worth to notice that the Eq.(\ref{B}) and the form
(\ref{Fc fin}) can be formally obtained if we write in
(\ref{int=Fá}) the current of the free system
(\ref{j=A mu g0}) instead of that of the interaction system and
change the covariant part of
(\ref{Amu}) for
\begin{equation}
A_\mu(s,Q^2,s')\left|_{P=p_c\;,P'=p'_c}\right.
= (p_c + p'_c\,)_\mu\;.
\label{A=P+P'}
\end{equation}

The Eq.(\ref{A=P+P'}) gives a simple prescription to write the
current matrix elements for interacting system in the basis
(\ref{PkJlSm}) in MIA using the current parameterization
(\ref{j=A mu g0}) for the free system.
The prescription is as follows: in the vectors in the
parameterization
(\ref{j=A mu g0}), (\ref{Amu}) one has to use the momenta of
composite system instead of the center--of--mass momenta of the
free two--particle system. Note that this prescription works for
more complicated systems, too.

We do not discuss in this paper the problem of going beyond the
limits of MIA and of obtaining corrections to
$g_0(s\,,Q^2\,,s')$ in (\ref{G=g0+Gc}), (\ref{Fc fin}).
This means that if considering, for example, nucleon systems we
do not take into account meson current.

Let us consider now the fulfilling of the conditions (i)--(vi)
for the electromagnetic current.

The conditions (i)--(iii) are satisfied by construction. For
example the fulfilling of (i) and (iii) is ensured by the
correct transformation properties of the 4-vectors in
(\ref{j=A mu g0}), (\ref{j=BG}), and (\ref{B}).

The condition (iv) is satisfied immediately as
the form factor $g_0(s\,,Q^2\,,s')$ in (\ref{j=A mu g0})
and the form factor $G(s\,,Q^2\,,s')$ in (\ref{j=BG})
are scalars in our simple model
\footnote{
The currents which do not conserve the parity also can be
considered in our formalism. In that case one can separate not
only the scalar part of current matrix element but the
pseudoscalar part, too. This case is considered elsewhere.
}.

The condition of cluster separability (v) needs a more detailed
consideration. At large distances (or if the interaction is
switched off) the contribution of two--particle currents has to
go to zero:
$G_c(s\,,Q^2\,,s') \to 0$  in (\ref{G=g0+Gc}).
This means that in the form (\ref{G=g0+Gc})
the form factor $G(s\,,Q^2\,,s')$ has to transform into
$g_0(s\,,Q^2\,,s')$.  Let us remark that the condition of
cluster separability is fulfilled in MIA, too, as in this
approximation the use of $g_0(s\,,Q^2\,,s')$ instead of
$G(s\,,Q^2\,,s')$ is supposed from the very beginning. When the
interaction is switched off the generalized function
$g_0(s\,,Q^2\,,s')$ for the free two--particle system acts on
a larger space of test functions than
(\ref{psi(ss')}). As $g_0(s\,,Q^2\,,s')$ contains only
the one--particle current contributions
(\ref{jP=jp1p2}) the condition (v)
is satisfied and the composite--system current go to the sum of
the one--particle currents.

The condition on the charge to be nonrenormalizable also is
fulfilled directly in MIA because the weak limit
(\ref{Q2=0}) does exist on test functions (\ref{psi(ss')}).

So, our prescription for the construction of the current in MIA
satisfies all the conditions for the current operator.

\subsection*{E. MIA {\large\it versus\ }IA}

Let us compare the approximation MIA with the well known IA.

To do this let us first calculate the form factor in IF RHD
not using the canonical parameterization. In particular, let us
formulate IA in terms of operators as it is formulated usually
(not in terms of form factors). Let us decompose the matrix
element (\ref{j=Fc}) through the complete set of states
(\ref{p1m1p2m2}):
$$
\langle\,p_c\,|j_\mu(0)|\,p_c'\,\rangle  =
\int\,\frac{d\vec p_1\,d\vec p_2} {2p_{10}\,2p_{20}}
\frac{d\vec p_1\,'\,d\vec p_2\,'}{2p_{10}'\,2p_{20}'}\,
$$
$$
\times \langle\,p_c\,|\vec p_1;\vec p_2\,\rangle
\langle\vec p_1;\vec p_2\,|j_\mu|\,\vec p_1\,';\vec p_2\,'\rangle
$$
\begin{equation}
\times \langle\,\vec p_1\,';\vec p_2\,'|p_c'\rangle\;.
\label{jc=j1+j2}
\end{equation}
Here $\langle\,\vec p_1;\vec p_2|p_c\rangle$ is wave function of constituents
in composite system.

If the current matrix element in ~(\ref{jc=j1+j2})
is taken in the IA approximation
~(\ref{j=jk+jkm}) and contains one--particle currents only, then
the Eq. ~(\ref{jc=j1+j2}) is selfcontradicting
~\cite{KeP91}.  In fact, one can show this in the following way.
In our simple model all the dynamical information about the
current (i.e. the composite system form factor) can be obtained
from only one matrix element in the Breit frame. However, to go
to the Breit frame one has to perform the transformation which
is interaction depending. This means that in Breit system
two--particle currents appear along with one--particle ones.
The form factors calculated in arbitrary coordinate frames using
different matrix elements will be of different forms.

To write the form factor in terms of wave functions
(\ref{wf}) one has to perform the CG decomposition of the basis
(\ref{p1m1p2m2}) in terms of the basis (\ref{PkJlSm})
in the wave functions (\ref{jc=j1+j2}) and to use the explicit
form for CG coefficients  (\ref{KGpr}) for the quantum numbers
of the system:
\begin{equation}
\langle\,\vec p_1;\vec p_2\,|p_c\rangle  = \sqrt{\frac{2}{\pi}}\,
\langle\,\vec P\,,\sqrt{s}\,,J\,,l\,,S\,,m_J\,|p_c\rangle .
\label{wfp1p2}
\end{equation}

The current matrix element in
(\ref{jc=j1+j2}) has the form (\ref{j=j1xI}).
The one--particle currents are expressed through the form
factors (\ref{j=f1}).

The Eq.(\ref{jc=j1+j2}) is an equality for two 4--vectors.
Taking different components of this equality and exploiting
$\delta$--functions in integrals, one can calculate the form
factor of the composite system. The result of calculation of the
form factor in this way is not unambiguous. In particular, it
depends on the actual choice of the component of the current
(\ref{jc=j1+j2})
to be used in the calculation.
Moreover, the result depends on the coordinate frame chosen to
perform the integration in
(\ref{jc=j1+j2}). This is the general feature of IA in the usual
formulation of IF RHD (see, e.g., \cite{KeP91}).

Let us write the final result of the calculation of the form
factor from the equation for the null--component of the current
and performing the integration in the coordinate frame where
$\vec p_c\,' = 0\>,\>\vec p_c = (0,0,p)$.
If now we write the integral in terms of the invariant variables
$s,s'$ the obtained form factor has the form:
$$
F_c(Q^2) =
\frac{M_c}{4}\frac{\sqrt{2\,(2\,M_c^2 + Q^2)}} {4\,M_c^2 + Q^2}\,
$$
$$
\times
\int\,\sqrt{\frac{s}{s'}}\,\frac{d\sqrt{s}\,
d\sqrt{s'}}{\sqrt{(s - 4\,M^2)(s' - 4\,M^2)}}
\frac{(s + s' + Q^2)^4\,Q^2}{[\lambda(s\,,-\,Q^2\,,s')]^{3/2}}
$$
\begin{equation}
\times
\frac{1}{(s\,s')^{1/4}}\frac{\theta(s,Q^2,s')}{\sqrt{s'}\,
(s + Q^2)}\,\varphi(s)\, \varphi(s')\,f_1(Q^2)\;.
\label{Fc conv}
\end{equation}
The Eq.(\ref{Fc conv}) differs from (\ref{Fc fin}),
obtained with the use of the two--particle free form factor. In
the case of wave functions satisfying the conditions
(\ref{phi(s)}), (\ref{norm}), the form factor (\ref{Fc conv})
satisfies the normalization: $F_c(0) = e_c$.
Let us note that the form factor obtained in this way from the
third current component in (\ref{jc=j1+j2}) does not satisfy
this condition.

Let us compare IA and MIA results.
Let us note once again that in MIA we separate (by use of the
scheme of canonical parameterization) the covariant part of the
current matrix element in
(\ref{int BG=Fá}) prior to perform any calculations. This
covariant part ensures the correct transformation properties of
the corresponding decompositions in terms of free--particle
states.
The difference between
(\ref{Fc fin}) and (\ref{Fc conv}) is:
$$
\Delta F_c(Q^2) =
\int\,d\sqrt{s}\,d\sqrt{s'}\,\,\varphi(s)\,\varphi(s')\,
$$
\begin{equation}
\times g_0(s,Q^2,s')\,\left[\,1 - R(s,Q^2,s')\right]\;.
\label{DFc}
\end{equation}
$$
R(s,Q^2,s') =
\frac{M_c}{2}\frac{\sqrt{2\,(2\,M_c^2 + Q^2)}} {4\,M_c^2 + Q^2}\,
\sqrt{\frac{s}{s'}}\,
$$
\begin{equation}
\times \frac{(s + s' + Q^2)^2}{(s\,s')^{1/4}}
\frac{1}{\sqrt{s'}\,(s + Q^2)}\;.
\label{R}
\end{equation}

The value $R(s,Q^2,s')$ presents an additional factor to
one--particle currents, that is in reality the two--particle
current contributions. This term ensures the Lorentz covariance
of the electromagnetic current matrix element and the current
conservation law in  (\ref{int dmu=Fá}). Let us note that this
additional term contains no dynamical information on the
interaction of test particle with two constituents
simultaneously. It does not depend, for example, on the
interaction constants for such a process.

So, to summarize, we can write the following schematic
equations:
$$
(IA)_{Breit} \ne (IA)_{Lab}
$$
$$
(MIA)_{Breit} = (MIA)_{Lab}
$$
It is well known that the standard IA depends strongly on the
coordinate frame used for the calculation. The MIA results
do not depend on it at all. So, the differences between IA and
MIA results for different IA coordinate frame can be
rather significant.

Notice that IA and MIA coincide in the nonrelativistic limit. As
this takes place, the nonrelativistic limits of form factors,
which were obtained from the different current components, are
identical. Hence the difference between IA and MIA is connected
with the breaking of relativistic covariance conditions really.

We give the quantitative comparison of the
form factors obtained in IA and in MIA in the Section IV
where the realistic calculation of the pion electromagnetic
structure is given.

\subsection*{F. The nonrelativistic limit}

The description of composite--system form factors in terms of
distributions is not a specific feature of our relativistic
approach. The similar formalism is used in nonrelativistic
theory of composite systems \cite{BrJ76} for a rather long
time (although not referring to the mathematics of
distributions). In the nonrelativistic limit our approach gives
the formalism developed in \cite{BrJ76}.

In the nonrelativistic limit the relativistic charge form factor
(\ref{Fc fin}) has the following form:
\begin{equation}
F_{NR}(Q^2) =
\int\,k^2\,dk\,k'\,^2\,dk'\,u(k)\,g_{0NR}(k,Q^2,k')\,u(k')\;,
\label{FNR}
\end{equation}
\begin{equation}
g_{0NR}(k,Q^2,k') = \frac{f_1(Q^2)}{k\,k'\,Q}\theta(k,Q^2,k')\;,
\label{g0NR}
\end{equation}
$$
\vartheta(k,Q^2,k') = \theta\left(k' - \left|k - \frac{Q}{2}\right|\right) -
\theta\left(k' - k - \frac{Q}{2}\right)\;.
$$
Here $g_{0NR}(k,Q^2,k')$ is the free relativistic form factor
obtained from (\ref{ff-nonint}) in the nonrelativistic limit.
$f_1(Q^2)$ is the charged--particle form factor.
The obtained result coincides with that derived in standard
nonrelativistic calculations \cite{BrJ76}.

In \cite{BrJ76} the same formulae are obtained from the
equations for form factors in terms of coordinate representation
wave functions:
\begin{equation}
F_{NR}(Q^2) =
f_1(Q^2)\,\int_0^\infty\,dr\,r\,u^2(r)\,j_0\left(\frac{Qr}{2}\right)\;.
\label{Fu(r)}
\end{equation}
The Eqs.(\ref{FNR}), (\ref{g0NR}) can be obtained from
(\ref{Fu(r)}) by use of the Bessel transformation:
\begin{equation}
u(k) =
\sqrt{\frac{2}{\pi}}\int_0^\infty\,r\,dr\,u(r)\,j_0\left(k\,r\right)
\end{equation}
and the normalization condition:
$$
\int_0^\infty\,u^2(r)\,dr = \int_0^\infty\,k^2\,u^2(k)\,dk = 1\;.
$$

Rigorously speaking, the Eq.(\ref{FNR})
has to be interpreted as a functional in the sense of
distributions generated by the function
$g_{0NR}(k,Q^2,k')$ and defined on  test functions
$u(k)\,u(k')$.  The ordinary function
(\ref{g0NR}) generates regular
generalized function defined generally on the larger
class of test functions
$\psi(k,k')$ in $R^2$, providing the uniform convergence of the
integral. One needs the uniform convergence to take limits
in the integrands.

Let us define the functional in $R^2$
by the following regular distribution (compare with
(\ref{<>})--(\ref{dmu})):
$$
\langle\,\tilde g_{0NR}(k,Q^2,k')\,,\,\psi(k,k')\rangle
$$
\begin{equation}
= \int\,d\mu (k,k')\,\tilde g_{0NR}(k,Q^2,k')\,\psi(k,k')\;,
\label{<g0NR>}
\end{equation}
$$
\tilde g_{0NR}(k,Q^2,k') = \vartheta(k)\,\vartheta(k')
g_{0NR}(k,Q^2,k')\;,
$$
$$
d\mu (k,k') = d\mu (k)\,d\mu (k')\>,\quad d\mu (k) = k^2\,dk\;.
$$

The function $g_{0NR}(k,Q^2,k')$ which appears in
\cite{BrJ76} quite formally, here has a definite physical
meaning and describes the electromagnetic properties of
nonrelativistic free system of two spinless particles in the
$S$ -- state, one of particle having no charge (compare with
$g_0(s,Q^2,s')$ in (\ref{j=A mu g0}), (\ref{ff-nonint}),
(\ref{<>})).  The static limit
$\lim_{Q^2\to 0}\,g_{0NR}(k,Q^2,k')$
giving the system charge exists only in the weak sense as the
limit of the functional (\ref{<g0NR>}):
$$
\lim_{Q^2\to 0}\,\langle\,\tilde g_{0NR}(k,Q^2,k')\,,\,\psi(k,k')\rangle
$$
\begin{equation}
= \langle\,e_c\,\delta(\mu(k') - \mu(k))\,\vartheta(k),\psi(k,k')\rangle\;.
\label{slimNR}
\end{equation}

On the test functions $\psi(k,k') = u(k)\,u(k')$ (with
$u(k)$ -- being the normalized bound state wave function),
the functional (\ref{<g0NR>}) defines the bound state form
factor in the nonrelativistic IA
(\ref{FNR}).  The weak limit (\ref{slimNR})
is equal to the system charge:
$$
\lim_{Q^2\to 0}\,\langle\,\tilde g_{0NR}(k,Q^2,k')\,,\,\psi(k,k')\rangle
$$
\begin{equation}
= e_c\,\int_0^\infty\,k^2\,dk\,u^2(k) = e_c\;.
\end{equation}
So, one can see that the description of the system in terms of
form factors in IA by the
Eq.(\ref{FNR}) (as in \cite{BrJ76}) in fact defines the form
factor in the sense of distributions as a functional defined on
the set of the wave functions of bound system.

To go beyond nonrelativistic IA one has to addend some terms
to $g_{0NR}(k,Q^2,k')$. For example, such terms cause the meson
exchange currents in two--nucleon systems. So, in the standard
nonrelativistic theory the dynamical treatment of exchange
currents is performed in the same way as in our relativistic
approach (\ref{G=g0+Gc}).

So, to conclude, one can consider our approach to IA to be a
relativistic generalization of nonrelativistic IA, and our
equations for form factors in this approximation to be a
relativistic generalization of the equations of
\cite{BrJ76}. Let us remark that in more complicated systems
(e.g., in $\rho$ -- meson and deuteron) our relativistic form factors
also have correct nonrelativistic limits which coincide with
\cite{BrJ76}.

\subsection* {G. A bridge to dispersion relations}

Let us discuss now one of the unsolved problems
of RHD -- the possible links between RHD and quantum field
theory (QFT) \cite{KeP91}. The fact, that RHD, contrary to QFT
itself, operates with the finite number of degrees of freedom,
makes it to be in some way similar to the dispersion approach of
QFT, which is dealing in principle with a finite number of
degrees of freedom, too. However, the dispersion relations,
based on the analytic properties of the scattering amplitudes,
matrix elements, form factors in the complex energy plane, are
rather correctly derived in the frame of QFT
\cite{Bar65}. So, it seems to us, that one can look for links
between RHD and QFT not only directly but through the dispersion
approach, too.

Here, using the simple model of the previous Subsections,
we compare our version of RHD with the so called modified
dispersion approach.  Dispersion-relation integrals over
composite-particle mass are used in this approach.
This approach enabled one to write the deuteron
electromagnetic form factors in terms of the physical hadron
scattering phase shift and gave the results for the elastic
$ed$--scattering in good agreement with experimental data.
The details of the modified dispersion approach can be found in
\cite{TrS69,ShT69,Tro94,MuT79}  (see also~\cite{AnK92}).  Let us
note that an immediate application of
the approach to quark systems is difficult to realize because of
the fact of quark confinement. However, there are some
investigations based on similar ideas where the form factors of
hadrons as constituent--quark bound states are considered in the
frame of the dispersion technique of the integral over composite
particle mass ~\cite{Mel94}.

For convenience of reader let us describe briefly, omitting the
proofs, the essence of the modified dispersion approach and
obtain the electromagnetic form factor of composite system for
our simple model following the paper \cite{TrS69}.

The Heisenberg current operator of the system of two particles
interacting as in ~(\ref{j=jk+jkm}) can be written in the form
\begin{equation}
j_\mu = j_\mu^{(0)} + j_\mu^{(int)}\;.
\label{j=j0+jc}
\end{equation}
Here $j_\mu^{(0)}$ is the current of the free two--particle
system (see (\ref{j=j1xI}), (\ref{j=A mu g0})).
The operator $j_\mu^{(int)}$  is interaction dependent.
Let us suppose that our model constituent system has scattering
states and let us calculate the matrix element
of the operator
(\ref{j=j0+jc}) between $in$-- and $out$-- states.
Let us suppose that the scattering states are the $S$--
states of relative motion. The matrix element of the operator
$j_\mu^{(0)}$ can be written in terms of the free two--particle
form factor (\ref{j=A mu g0}) calculated previously (\ref{ff-nonint}).
The matrix element of the operator containing the interaction
$$
\langle\vec P(\pm)\mid j_\mu^{(int)}\mid \vec P'(\pm)\,\rangle
$$
can be written in terms of the form factor
\begin{equation}
G_i(s\mp i\varepsilon,Q^2,s'\pm i\varepsilon)\;.
\label{jc=a mu Gc}
\end{equation}
Here sign "$+$" stands for $in$-- state and sign "$-$" for
$out$-- state. The form factor in (\ref{jc=a mu Gc}) has
kinematic cuts in the complex plane of the variables $s\;,\;s'$.
The cuts go along the real axis from the point $4\,M^2$ to
infinity. The notation $G_i(s+i\varepsilon,Q^2,s'-i\varepsilon)$
means that there exists the analytic continuation of this
form factor from the physical region of the variable $s$ into
the upper complex half--plane, and to the lower half--plane in
$s'$. One can check this fact considering simple models.
The form factor entering the parameterization of the total
current (\ref{j=j0+jc}) can be written as the sum of the form
factors:
\begin{equation}
G(s,Q^2,s') = g_0(s,Q^2,s') + G_i(s,Q^2,s')\;.
\label{G=Gi+g0}
\end{equation}
Let us consider the matrix element of the total current. Let us
fix the variable $s$ and let us connect by $S$--matrix the
$in$-- and $out$-- vectors of the basis in the variable $s'$:
\begin{equation}
\langle\vec P\mid j_\mu\mid \vec P'(+)\,\rangle =
\langle\vec P\mid j_\mu\mid \vec P'(-)\,\rangle \,S(s')\;.
\label{+=S-}
\end{equation}
$S(s) = \exp(2\,i\,\delta)$, $\delta$
is the scattering phase shift. $S$ -- matrix can be written in
the form:
\begin{equation}
S(s) = \frac{B(s - i\varepsilon)}{B(s + i\varepsilon)}\;.
\label{S=B/B}
\end{equation}
Here $B(s)$ is the relativistic analog of the Jost function.
Taking into account Eqs.(\ref{jc=a mu Gc}), (\ref{S=B/B})
one can rewrite (\ref{+=S-}) as:
$$
G_i(s,Q^2,s'- i\varepsilon)\,B(s' - i\varepsilon) -
G_i(s,Q^2,s'+ i\varepsilon)\,B(s' + i\varepsilon)
$$
\begin{equation}
= -\,g_0(s,Q^2,s')\,(B(s' - i\varepsilon) - B(s' + i\varepsilon))\;.
\label{RiG}
\end{equation}
The equation (\ref{RiG}) presents the so called Riemann--Hilbert
problem for half--axis. The solution has the form
\cite{Gak66}:
\begin{equation}
G_i(s,Q^2,s')\,B(s') =
\tilde G(s,Q^2,s') + C_1(s,Q^2,s')\;,
\label{Gc=G+C1}
\end{equation}
\begin{equation}
\tilde G(s,Q^2,s') =
-\,\frac{1}{2\,\pi\,i}\int_{4\,M^2}^\infty\,
\frac{ds''\,g_0(s,Q^2,s'')\,\Delta(s'')}{s'-s''}\;,
\label{G=int}
\end{equation}
$$
\Delta(s) = (B(s + i\varepsilon) - B(s - i\varepsilon))\;,
$$
Here $C_1(s,Q^2,s')$  is an unknown function, regular in $s'$ in
the neighbourhood of the real axis for
$4\,M^2\>\le\>s'\><\infty$.
Now let us connect the $in$ --  and $out$ -- bases in the
variable $s$. Taking into account the explicit form of
$G_i$ (\ref{Gc=G+C1}), (\ref{G=int}) we obtain the boundary
value Riemann--Hilbert problem for the function
$C_1$ in the variable $s$ with the solution:
$$
C_1(s - i\varepsilon,Q^2,s')\,B(s - i\varepsilon) -
C_1(s + i\varepsilon,Q^2,s')\,B(s + i\varepsilon)
$$
\begin{equation}
= \left[\,g_0(s,Q^2,s')\,B(s') + \tilde
G(s,Q^2,s')\right]\,\Delta(s)\;,
\label{RiGC1}
\end{equation}
$$
B(s)\,C_1(s,Q^2,s') =
-\,\frac{1}{2\,\pi\,i}\int_{4\,M^2}^\infty\,ds'''
$$
$$
\times \frac{\left[g_0(s''',Q^2,s')\,B(s') +
\tilde G(s''',Q^2,s')\right] \Delta(s''')}{s-s'''}
$$
\begin{equation}
+ C(s,Q^2,s')\;.
\label{C1=int}
\end{equation}
The unknown function $C(s,Q^2,s')$ is regular in the
neighbourhood of the real axis for
$4\,M^2\>\le\>s,s'\><\infty$ in $s'$ as well as in $s$.
Now let us consider the matrix element of the total current
(\ref{j=j0+jc}), (\ref{G=Gi+g0}) between $in$--states.
Substituting (\ref{Gc=G+C1}), (\ref{G=int}), (\ref{C1=int}) in
(\ref{G=Gi+g0}) we obtain finally the following form for the
form factor of the total current (\ref{j=j0+jc}) in $in$--basis:
$$
G(s,Q^2,s') = g_0(s,Q^2,s')
$$
$$
- \frac{1}{2\,\pi\,i\,B(s'+ i\,\varepsilon)}
\int_{4\,M^2}^\infty\,ds''
\frac{g_0(s,Q^2,s'')\,\Delta(s'')}{s'-s'' + i\,\varepsilon}
$$
$$
- \frac{1}{2\,\pi\,i\,B(s - i\,\varepsilon)}
\int_{4\,M^2}^\infty\,ds'''
\frac{g_0(s''',Q^2,s')\,\Delta(s''')}{s - s''' - i\,\varepsilon}
$$
$$
- \frac{1}{4\,\pi^2\,B(s - i\,\varepsilon)\,B(s' + i\,\varepsilon)}
$$
$$
\times \int_{4\,M^2}^\infty\,ds'''\,\int_{4\,M^2}^\infty\,ds''
\frac{g_0(s''',Q^2,s'')\,\Delta(s''')\,\Delta(s'')}
{(s - s''' - i\,\varepsilon)(s' - s'' + i\,\varepsilon)}
$$
\begin{equation}
+ \frac{C(s,Q^2,s')}{B(s - i\,\varepsilon)B(s' + i\,\varepsilon)}\;.
\label{Gfin}
\end{equation}
The Eq.(\ref{Gfin}) provides the correct analytic properties of
the form factor obtained in QFT approach
\cite{Bar65}. In particular, this form contains the anomal
branch points known from the dispersion approach to composite
systems (e.g., to the deuteron). The Eq.(\ref{Gfin})
can be used to obtain the form factor of the constituent bound
state for the case of the $S$--state of relative motion.
Now it is necessary to perform the analytic continuation of
(\ref{Gfin}) in the variables $s,s'$ to the bound state point
$s = s' = M_c^2$ ($M_c$ is the bound state mass) and to take
the residues in the poles.
As the result we obtain the bound--system form factor directly
in terms of the $S$--scattering phase shift for constituents:
\begin{equation}
 F_c(Q^2) =
\Gamma^2\,\int_{4\,M^2}^\infty\,ds\,ds'
\frac{g_0(s,Q^2,s')\,\Delta(s)\,\Delta(s')}
{(s - M_c^2)(s' - M_c^2)}\;.
\label{Fc dr}
\end{equation}
Now the constant $\Gamma^2$ is determined by the condition
$F_c(0)=e_c$ and indirectly takes into account the contributions
of the so called unphysical cuts. The Jost--function
discontinuities can be written in terms of experimental
scattering phase shift. The free two--particle form factor for
our model has the form (\ref{ff-nonint}).  $F_c(Q^2)$ is the
functional generated by the generalized function
$g_0(s,Q^2,s')$ on the test functions $\Delta(s)/(s - M_c^2)$.
The described formalism was applied to the deuteron in Ref.
\cite{MuT79} and gave a good agreement with experimental data.

Let us note, that the form (\ref{Fc dr})
obtained through the modified dispersion approach is in close
analogy to the forms
(\ref{Fc fin}), (\ref{FNR}) obtained in the frame of IF RHD.
This analogy can be made even more obvious using the results of
the Ref. ~\cite{Tro94} where
neutron-proton system was considered in nonrelativistic case.
In ~\cite{Tro94} it is shown  that if the deuteron
electrodisintegration amplitude satisfies Mandelstam
representation, then the wave function of the system has the
well fixed form and can be expressed in terms of $np$
-scattering phases. This wave function satisfies a dispersion
relation.  The analytical properties used during the derivation
of this relation are the same for the large group of
phenomenological potentials. So the obtained dispersion relation
can be used to find explicit form for real two-nucleon systems
wave functions. As one can see from the solution structure, such
a reconstruction of wave function is stable both in the usual
sense and in the sense of the large energy phases influence.
Finally, the bound state wave function $u(r)$ is of the form:
\begin{equation}
u(r) =
\tilde \Gamma\,\int_{-\infty}^\infty\,dx
\frac{\Delta(x)} {(x - \kappa)} [\sin (xr)] .
\label{u(r)}
\end{equation}
(See ~\cite{Tro94} for details.)
Here $\kappa ^2$ is the deuteron binding energy and the
nonphysical cuts contribution enters the normalization constant
$\tilde \Gamma$.

In nonrelativistic case the Jost--function discontinuity
multiplied by the pole term in
(\ref{Fc dr})  gives the nonrelativistic wave function up to
nonphysical cuts contribution ~\cite{Tro94}.  When these
contributions are taken into account the equations
(\ref{FNR}) and (\ref{Fc dr}) do coincide.

So, the relativistic Eq.(\ref{Fc fin}) can be motivated
(at least for composite
systems which have the scattering states) in the frame of
modified dispersion approach, that is on the usual level of
correctness for obtaining the analytical properties of the form
factors of composite systems in the frame of QFT.


\section{The electroweak structure of pion}

Now we apply the method of previous sections to the calculation
of the electroweak structure of pion. There exists a lot of
experimental data on pion, so the effectiveness of the method can
be checked by the comparison with the data (see, e.g., \cite{Jau91} and
references therein).

\subsection*{A. The electromagnetic form factor of pion}

There are many facts that make it interesting to consider the
pion in the frame of the formalism developed in the previous
sections. First, the pion is an important object in the particle
physics and is in the focus of interest for years. Second, the
pion consists of light quarks and thus has to be considered in
the frame of relativistic approach. Third, at the present time
the large program on the pion structure is on line in JLab
\cite{CEBAF93}.

We consider the pion as a system of two constituent quarks. The
system is described by a phenomenological wave function.

In theoretical treatment of the pion electromagnetic structure
one has to make difference between "soft" and "hard" parts of
the form factor. The "soft" part which dominates at small and
intermediate momentum transfers, needs nonperturbative
approaches. The "hard" part which defines the form factor at
asymptotically large values of momentum transfers can be
calculated from perturbative QCD. However, a controversy still
exists concerning the scale of momentum transfers characteristic
of the transition from the nonperturbative to the perturbative
regime (see, e.g. ~\cite{IsL8489}).  It is pointed out by
different authors that the existing experimental data are
defined by the "soft" part of the form factor and can be
described by use of phenomenological wave functions of
constituent quarks without involving the perturbative QCD.
Usually such calculations were performed in the frame of
light--front relativistic quantum mechanics. In the present
paper we calculate the "soft" part of the charge form factor of
pion in the frame of IF RHD. We obtain a good description of the
behaviour of the form factor in the wide region of momentum
transfers where the experimental data exist
$0\;\leq \;Q^2\;\leq 8(GeV/c)^2$ ~\cite{BaK96}.

We pay a special attention to the role of relativistic
properties of quark spins for the pion structure. The
relativistic spin rotation effect (Wigner rotation)
caused by the summation of quark spins gives large
contribution to the pion form factor ( 10 -- 20\%
depending on the value of momentum transfer)
~\cite{KrT99}. It is interesting that spin rotation effect
vanishes as $Q^2\; \to \;0$ but gives large contribution
($\sim 30\%$) to the pion charge radius
~\cite{KrT93}, which is defined by the slope of the form factor
at $Q^2=0$.

In our version of IF RHD the meson form factor asymptotics
at large momentum transfer is the same as in perturbative QCD.
The asymptotics is now determined by relativistic kinematics
only, specifically by the relativistic effect of spin rotation,
and does not depend on the choice of the quark wave function,
that is of the quark interaction model.

Our version of IF RHD approach gives, in agreement with the
experimental data, an adequate description of the pion
electromagnetic structure in large region of momentum transfers.

The pion is spinless, so the electromagnetic current matrix
element has the form
(\ref{j=Fc}) with  $p_c\to p_\pi\;,\;F_c(Q^2)\to F_\pi(Q^2)$. In the frame of
composite quark model pion is considered as the bound state of
$u$-- and $\bar d$-- quarks. We suppose that quark masses are
equal: $m_u = m_d = M$.

To calculate in MIA the composite--system form factor one needs
to construct first the free two--particle form factor
(\ref{j=A mu g0}), (\ref{ff-nonint}), (\ref{Fc fin}).
Contrary to the simple model of the previous Section now we
consider the system of two charged particles with spins 1/2.
This gives the following complications. First, the Eq.
(\ref{j=j1xI}) for the current operator of the free system is
now transformed to the form:
\begin{equation}
j^{(0)}_\mu(0) =
j_{1\mu}\otimes I_{2}\oplus j_{2\mu}\otimes I_{1}\;.
\label{j=j1xI+j2xI}
\end{equation}
Here $j_{(1,2)\mu}$ - the electromagnetic currents of
particles,
$I_{(1,2)}$ -- the unity operators in the one--particle state
Hilbert spaces.
The Eq.(\ref{j=j1xI+j2xI}) can be rewritten in terms of matrix
elements:
$$
\langle\vec p_1,m_1;\vec p_2,m_2|j_\mu^{(0)}(0)|
\vec p\,'_1,m'_1;\vec p\,'_2,m'_2\rangle  =
$$
\begin{equation}
= \langle\vec p_2,m_2|\vec p\,'_2,m'_2\rangle
\langle\vec p_1,m_1|j_{1\mu}|\vec p\,'_1,m'_1\rangle
+ (1\leftrightarrow 2)\;.
\label{j=j1+j2}
\end{equation}

Second, the matrix element of one--particle current contains
now, contrary to (\ref{j=f1}), the magnetic form factors of
quarks as well as the charge ones.
Now the parameterization (the elementary--particle one following
\cite{KeP91}) is of the form:
$$
\langle\vec p,\,m|j^\mu(0)|\vec p\,',\,m'\rangle
$$
\begin{equation}
= \overline u_{\vec p\,m}\gamma ^\mu u_{\vec p\,'m'}\,F_1(Q^2) -
\overline u_{\vec p\,m}\sigma ^{\mu \nu}q_\nu \,u_{\vec p\,'m'}\,F_2(Q^2)\;,
\label{j=F1F2}
\end{equation}
$u_{\vec pm}$ - the Dirac bispinor,
$\gamma^\mu$ - Dirac matrix,
$$
\sigma^{\mu \nu} =
\frac{1}{2}(\gamma^\mu\gamma^\nu - \gamma^\nu\gamma^\mu)\;,\quad
q_\nu = (p - p')_\nu\;,
$$
Using multipole parameterization we can write the one--particle
current matrix element in terms of Sachs form factors:
$$
G_E(Q^2) =
\tilde F_1(Q^2) + \frac {\kappa Q^2}{4M^2}\,\tilde F_2(Q^2)\;,
$$
$$
G_M(Q^2) = \tilde F_1(Q^2) + \kappa \tilde F_2(Q^2)\;,
$$
\begin{equation}
F_1(Q^2) = e\tilde F_1(Q^2)\;,\quad
F_2(t) = \frac {\kappa }{2M} \tilde F_2(Q^2)\;.
\label{G=F}
\end{equation}
Here $G_{E,M}$ - Sachs electric and magnetic form factors,
respectively,
$e$ is the particle charge,
$\kappa $ is the anomalous magnetic moment.

It is convenient to use the canonical parameterization of matrix
elements \cite{ChS63}:
$$
\langle\,\vec p,\,m\,|\,j_\mu(0)\,|\,\vec p\,',\,m'\,\rangle
$$
$$
= \sum_{m''}\,\langle m|D^j(p,\,p')|m''\rangle
\langle m''|\,f_1(Q^2)K'_\mu + i\,f_2(Q^2)R_\mu|m'\rangle\;,
$$
\begin{equation}
K'_\mu = (p + p')_\mu \,, \quad
R_\mu = \epsilon _{\mu \,\nu \,\lambda \,\rho}\, p^\nu
\,p'\,^\lambda \,\Gamma^\rho (p')\;.
\label{j=f1f2}
\end{equation}
$\Gamma(p)$ is 4--vector of spin:
$$
\vec \Gamma(p) =  M\,\vec j + \frac {\vec p(\vec p\vec j)}{p_0 + M}\;,\quad
\Gamma_0(p) = (\vec p\vec j)\;.
$$
The form factors
$f_1(Q^2)$ ¨ $f_2(Q^2)$ are the electric and magnetic form
factors of particles. They are connected with Sachs form
factors \cite{BaY95}:
$$
f_1(Q^2) = \frac {2M}{\sqrt {4M^2 + Q^2}}\,G_E(Q^2)\;,
$$
\begin{equation}
f_2(Q^2) = -\frac {4}{M\sqrt {4M^2 + Q^2}}\,G_M(Q^2)\;.
\label{Bal}
\end{equation}

Third, now the CG coefficients are of more complicated form.
They are given by
(\ref{Klebsh}) with $J = S = l = 0$. Contrary to the previous
simple case, now the CG coefficients contain the Wigner rotation
matrices.

Finally, the free two--particle form factor for the system of
two particles with spin 1/2 and quantum numbers
$J = S = l = 0$ is of the form (see also \cite{BaK96}):
$$
g^{q\bar{q}}_0(s,Q^2,s')=
n_c\,\frac{(s+s'+Q^2)Q^2}{2\sqrt{(s-4M^2) (s'-4M^2)}}\;
$$
$$
\times \frac{\theta(s,Q^2,s')}{{[\lambda(s,-Q^2,s')]}^{3/2}}
\frac{1}{\sqrt{1+Q^2/4M^2}}
$$
$$
\times \left\{(s+s'+Q^2)(G^u_E(Q^2)+G^{\bar d} _E(Q^2))\cos(\omega_1+\omega_2)
+\right.
$$
\begin{equation}
+ \left.\frac{1}{M}\xi(s,Q^2,s')
(G^u_M(Q^2)+G^{\bar d}_M(Q^2)) \sin(\omega_1 + \omega_2)\right\}\;,
\label{g_0}
\end{equation}
Here
$$
\xi(s,Q^2,s')=\sqrt{ss'Q^2-M^2\lambda(s,-Q^2,s')}\;,
$$
$n_c$ is the number of quark colours,
$\omega_1$ ¨ $\omega_2$ -- the Wigner rotation parameters:
$$
\omega_1 =
\arctan\frac{\xi(s,Q^2,s')}{M\left[(\sqrt{s}+\sqrt{s'})^2
+ Q^2\right] + \sqrt{ss'}(\sqrt{s} +\sqrt{s'})}\;,
$$
\begin{equation}
\omega_2 = \arctan\frac{
\alpha (s,s') \xi(s,Q^2,s')} {M(s + s' + Q^2)
\alpha (s,s') + \sqrt{ss'}(4M^2 + Q^2)}\;,
\label{omega}
\end{equation}
with $\alpha (s,s') = 2M + \sqrt{s} + \sqrt{s'}$, and
$G^{u,\bar d}_{E,M}(Q^2)$ are Sachs form factors for quarks.
The $\theta$ -- function in (\ref{g_0}) is the same as in (\ref{ff-nonint}).

An interesting effect follows from
(\ref{g_0}): due to the relativistic Wigner spin rotation
effect the pion charge form factor contains the contribution of
quark magnetic form factors.

The pion charge form factor can be calculated using
(\ref{Fc fin}), with
(\ref{g_0}) for the free two--particle form factor:
\begin{equation}
F_\pi(Q^2) = \int\,d\sqrt{s}\,d\sqrt{s'}\,
\varphi(s)\,g^{q\bar{q}}_0(s\,,Q^2\,,s') \varphi(s') .
\label{Fpi}
\end{equation}

\subsection*{B. The lepton decay constant of pion}

Let us calculate now the lepton decay constant of pion in the
frame of our approach. The interest to such a calculation is
threefold. First, this constant is measured in experiment with
great accuracy \cite{PDG},  so that it can be a test for the
model, and give the limits for parameters of models. Second, it
is interesting to describe the electromagnetic form factors and
the weak decay constant in the frame of one and the same
approach: the decay constant indirectly, through the parameters
of the model, defines the behavior of form factors at large
values of momentum transfers $Q^2$. Third, it is interesting to
estimate relativistic effects in the lepton decay.

The lepton decay constant $f_\pi$ is defined by the
electroweak--current matrix element
\cite{Jau91}:
\begin{equation}
\langle0|j_\mu(0)|\,p_\pi\,\rangle  =
if_\pi\,p_\pi\,_\mu\frac{1}{(2\pi)^{3/2}}\;.
\label{j=f_c}
\end{equation}
$p_\pi$ -- 4-momentum of meson. Let us decompose the l.h.s.
of (\ref{j=f_c}) in the basis (\ref{PkJlSm}). Using the explicit
form of the meson wave function
(\ref{wf}) one can obtain for (\ref{j=f_c}):
$$
\int\,\frac{N_c}{N_{CG}}\,d\sqrt{s}\,
\langle 0|j_\mu(0)|\vec p_\pi\,,\sqrt{s}\rangle \varphi(s)
$$
\begin{equation}
= if_\pi\,p_\pi\,_\mu\frac{1}{(2\pi)^{3/2}}\;.
\label{int ds=f_c}
\end{equation}
As in Section II (Eq.(\ref{j=BG})) one can divide the integrand
in (\ref{int ds=f_c}) into two parts:
the covariant part (smooth ordinary function) and the invariant
part.
\begin{equation}
\frac{N_c}{N_{CG}}\, \langle0|j_\mu(0)|\vec p_\pi\,,\sqrt{s}\rangle
= iG(s)B_\mu(s)\frac{1}{(2\pi)^{3/2}}\;.
\label{j=G(s)}
\end{equation}

The invariant form factor $G(s)$ is a generalized function. In
the same way as in calculating
(\ref{int G=Fá}) of the previous section, we now obtain the
lepton decay constant of pion in the form
\begin{equation}
\int\,d\sqrt{s}\,G(s)\varphi(s) = f_\pi\;.
\label{int G(s)=f_c}
\end{equation}

In general, the form factor
$G(s)$ can be calculated in the frame of the standard model for
electroweak interactions. However, in this paper we limit
ourselves by 4-fermion interaction. We take for
$G(s)$  the form factor which parameterizes the decay of free
two--quark system:
\begin{equation}
\langle0|j^{(0)}_\mu(0)|\vec P\,,\sqrt{s}\rangle  =
iG_0(s)P_\mu\frac{1}{(2\pi)^{3/2}}\;.
\label{j0=G0}
\end{equation}
The explicit form
(\ref{j0=G0}) is written by analogy to
(\ref{j=A mu g0})   not taking into account the current
conservation law. The form
(\ref{j0=G0}) is quite similar to
(\ref{j=f_c}) but instead of the constant
$f_\pi$ the form factor depending on invariant variables
is written.
To calculate $G_0(s)$ let us decompose (\ref{j0=G0}) in the
one--particle basis (\ref{p1m1p2m2}).  Now we obtain for
(\ref{j0=G0}):
$$
iG_0(s)P_\mu\frac{1}{(2\pi)^{3/2}}
$$
$$
= \sum_{m_1\,,m_2\,,i_c}\int\,\frac{d\vec p_1}{2p_{10}}\,
\frac{d\vec p_2}{2p_{20}}\,
\langle 0|j^{(0)}_{\mu\,i_c}|\vec p_1\,,m_1\,;\vec p_2\,,m_2\rangle
$$
\begin{equation}
\times \langle\vec p_1\,,m_1\,;\vec p_2\,,m_2|\vec P\,,\sqrt{s}\rangle .
\label{G0=int}
\end{equation}
$i_c =1,2,3$, the sum over $i_c$ is the sum over the colours.
The CG coefficients are known (\ref{Klebsh}).
The current matrix element in the basis
(\ref{p1m1p2m2}) can be written in the standard way in terms of
the lepton decay current matrix element
\cite{Jau91}:
$$
\langle 0|j^{(0)}_\mu|\vec p_1\,,m_1\,;\vec p_2\,,m_2\rangle
$$
\begin{equation}
=
\frac{1}{(2\pi)^{3}}\,\bar v(\vec p_2\,,m_2)
\gamma_\mu(1 + \gamma^5) u(\vec p_1\,,m_1)\;.
\label{j0=vgamu}
\end{equation}
We integrate in (\ref{G0=int}) in the coordinate frame with
$\vec P =0$.  Finally, we obtain:
\begin{equation}
G_0(s) =
\frac{n_c}{2\sqrt{2}\,\pi\,P_0} (p_{0}+M)\,
\left[1 - \frac{k^2}{(p_{0}+M)^2}\right]\;,
\label{G0}
\end{equation}
$$
p_{0} = \sqrt{k^2 + M^2}\;.
$$

Substituting (\ref{G0})
in the Eq.(\ref{int G(s)=f_c}) we obtain the result which has
the following form if written in invariant variables:
\begin{equation}
f_\pi =
\frac{2M\,n_c}{2\sqrt{2}\,\pi}\,\int\,d\sqrt{s}
\frac{1}{\sqrt{s}}\, \varphi(s)\;.
\label{fpi}
\end{equation}

Let us notice that the Eq.(\ref{fpi}) coincides with that
obtained in the frame of light--front dynamics
\cite{Jau91}. However, although all forms of RHD are unitary
equivalent
\cite{Lev95}, nevertheless after the physical approximations
are made in more complicated cases
the results, e.g. for form factors, can be
different.  This is possibly due to the fact that the unitary
operators connecting different forms of RHD are interaction
dependent \cite{Lev95} and so the RHD forms realize one and the
same approximation in different ways.

Let us remark that the nonrelativistic limit of the
Eq.(\ref{fpi}) gives the standard form in terms of coordinate
space wave function at zero value.


\subsection*{C. The results of calculations}

To calculate the electroweak structure of pion using
(\ref{Fpi}), (\ref{g_0}), (\ref{fpi}), (\ref{phi(s)})
the following meson wave functions were utilized:

1. A gaussian or harmonic oscillator (HO)  wave function
\begin{equation}
u(k) = N_{HO}\,\hbox{exp}\left(-{k^2}/{2b^2}\right).
\label{HO-wf}
\end{equation}

2. A power-law (PL)  wave function
\begin{equation}
u(k) = N_{PL}\,{(k^2/b^2 + 1)^{-n}}\>,\quad n = 2\>,3\;.
\label{PL-wf}
\end{equation}

3. The wave function with linear confinement from
Ref.\cite{Tez91}:
$$
u(r) = N_T \,\exp(-\alpha r^{3/2} - \beta r)\>,\quad
\alpha =\frac{2}{3}\sqrt{M\,a}\>,\quad
$$
\begin{equation}
\beta = \frac{M}{2}\,b\;.
\label{Tez91-wf}
\end{equation}
$a\>,b$ -- parameters of linear and Coulomb parts of potential
respectively.

In the Ref.\cite{BaK96} in the calculation of pion
electromagnetic structure we supposed the quarks to be
point--like. The results of \cite{BaK96} can be considered as
preliminary results. However, one has to take into account the
structure of constituent quarks \cite{Ger8995}, in particular,
the anomalous magnetic moment. As anomalous magnetic moments
are connected with finite size of quark, one has to take into
account the explicit form of quark form factors entering
(\ref{g_0}) and the pion charge form factor
(\ref{Fpi}).
As in \cite{CaG96} let us use the following forms for quark form
factors:
$$
G^{q}_{E}(Q^2) = e_q\,f(Q^2)\;,
$$
\begin{equation}
G^{q}_{M}(Q^2) = (e_q + \kappa_q)\,f(Q^2)\;.
\label{q ff}
\end{equation}
Here $e_q$ -- the quark charge, $\kappa_q$ -- the quark
anomalous magnetic moment (in natural units). To obtain the
explicit form of the function
$f(Q^2)$ let us consider the asymptotics of pion charge form
factor as
$Q^2\;\to\;\infty\;,\;M\;\to\;$0.

To obtain the asymptotic behavior let us first make the
asymptotic estimation of the integrals in (\ref{Fpi}) in the
point--like quark approximation
($f(Q^2) = 1\;,\;\kappa = 0$ in(\ref{q ff}) ).
Omitting the details of calculation (given in
\cite{KrT98}) we write the final result for the asymptotics in
the form:
\begin{equation}
F_\pi(Q^2)\quad\sim\quad Q^{-2}\;.
\label{as-qcd}
\end{equation}
The asymptotics does not depend on the actual form of the wave
function and coincides with that obtained in QCD. The actual
form we obtain, e.g. for
(\ref{HO-wf}) is:
\begin{equation}
F_\pi(Q^2)\>\sim 32\sqrt{2}
\frac{\left[\Gamma\left(\frac{5}{4}\right)\right]^2}{\sqrt{\pi}}
\frac{b^2}{Q^2}\;.
\label{as s pov}
\end{equation}
It is worth to compare the form
(\ref{as s pov}) with the detailed QCD result
\cite{Rad98}:
\begin{equation}
F_\pi(Q^2) = \frac{8\,\pi\,\alpha_s\,f_\pi^2}{Q^2}\;.
\label{Fpi qcd}
\end{equation}

If $\alpha_s/\pi\;\sim\;0.1$
then (\ref{as s pov}) and (\ref{Fpi qcd}) coincide at $b\;\sim\;$0.1.
So the asymptotics (\ref{as-qcd}) is quite realistic.

In the case of non--point--like quarks we obtain another
asymptotics because the form factor depends upon the momentum
transfer. It is known that QCD gives logarithmic corrections
to (\ref{Fpi qcd}). To agree with this QCD corrected
asymptotics we can, for example, choose the following form
for $f(Q^2)$:
\begin{equation}
f_q(Q^2) = \frac{1}{1 + \ln(1+ \langle r^2_q\rangle Q^2/6)}\;.
\label{f_qour}
\end{equation}
Here $\langle r^2_q\rangle$ is the MSR of
the constituent quark which can be considered as the model parameter. Let us
fix it (as in \cite{CaG96}) to be:
$\langle r^2_q\rangle \simeq 0.3/M^2$.

For the constituent quark mass in pion we use the value which is
usually used in the calculations in RHD: $M = $ 0.25 GeV.

The quark anomalous magnetic moments can be taken from
\cite{Ger8995}:  $\kappa_u = 0.029\;,\;\kappa_d = -\,0.059$.

We choose the parameters
$b$  in (\ref{HO-wf}), (\ref{PL-wf}) and $a$ in (\ref{Tez91-wf})
in such a way as to fit the pion MSR:
$\langle r_\pi^2\rangle  = (\hbox{0.432}\pm\hbox{0.016 ) Fm}^2\>$
\cite{Ame84}.
We choose this way to fix the model parameters because the pion
MSR is defined by the form factor at small values of
$Q^2$, that is the range where potential models work well.

The fit of the pion MSR gives the following parameters of the
wave functions: in the model (\ref{HO-wf})
$b$ = 0.2784 GeV, model (\ref{PL-wf}) at $n$ = 2 $b$ = 0.3394 GeV,
model (\ref{PL-wf}) at $n$ = 3 $b$ = 0.5150 GeV,  model (\ref{Tez91-wf})
$b = (4/3)\alpha_s$, $\alpha_s$ = 0.59  at light meson mass scale, $a$ =
0.0567 GeV$^2$.

The results of calculation are presented on Figs.3 and 4.

\begin{figure*}
\centerline{\epsfxsize=0.4\textwidth \epsfbox{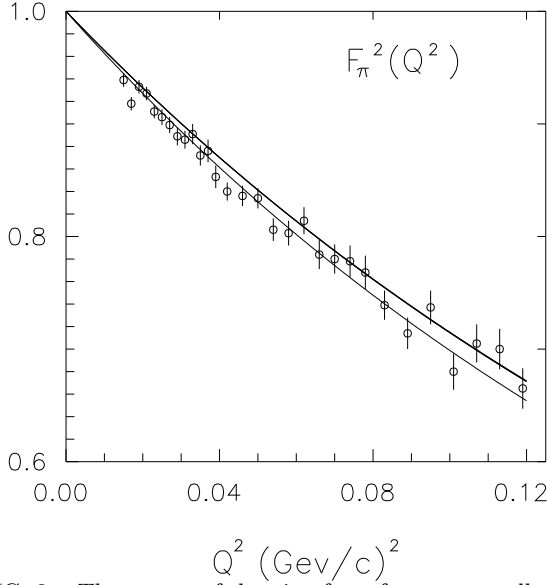}}
\caption{
The square of the pion form factor at small values of momentum
transfers for different models.}
\end{figure*}
The square of the pion form factor at small values of momentum
transfers for different models
(\ref{HO-wf}) -- (\ref{Tez91-wf}) is presented on Fig.3.
Results of
calculation in the models (\ref{HO-wf}), (\ref{PL-wf}) at $n$ = 3 and
(\ref{Tez91-wf})  coincide very closely.

The calculations of product $Q^2\,F_\pi(Q^2)$ at high
momentum transfers
for different models
(\ref{HO-wf}) -- (\ref{Tez91-wf}) are presented on Fig.4.
Legend is following: 1 -- harmonic oscillator wave function
(\ref{HO-wf}), 2 -- power--law wave function  (\ref{PL-wf}) at $n$ = 2,
3 -- power--law wave function  (\ref{PL-wf}) at $n$ = 3, and wave
function from
model with linear confinement (\ref{Tez91-wf}) (these curves coincide very
closely).

All the models for the interaction
(\ref{HO-wf}),
(\ref{PL-wf}), (\ref{Tez91-wf})
give a good description of the existing experimental data
\footnote{TheJLab new results \cite{Vol00} are discussed in
connection with our approach in \cite{KrT00a}}.

The dependence of the results on the actual model is much less
pronounced that in the case of point--like quarks \cite{BaK96}.

The lepton decay constants calculated following Eq.
(\ref{fpi}) with different wave functions have the following
values:
$f_\pi =0.1210$ GeV  in the model (\ref{HO-wf}), $f_\pi =0.1327$
GeV in the model (\ref{PL-wf}) with $n=2$,
$f_\pi =0.1282$ GeV in the model (\ref{PL-wf}) with $n=3$,
and $f_\pi =0.1290$ GeV in the model (\ref{Tez91-wf}).
Let us emphasize that we have used no fitting parameters to
calculate the lepton decay constant. Nevertheless, the obtained
values are very close to the experimental value:
$f_{\pi\>exp} = 0.1307\pm 0.0005$ GeV \cite{PDG}.

\begin{figure*}
\centerline{\epsfxsize=0.4\textwidth \epsfbox{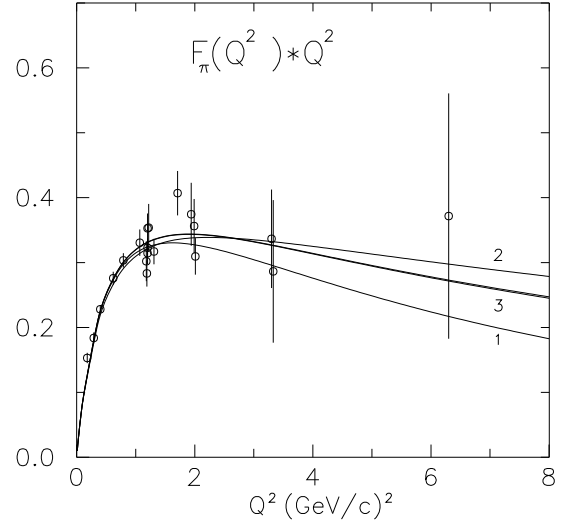}}
\caption{
Electromagnetic form factor,\protect$Q^2\,F_\pi(Q^2)$, at high
momentum transfers.}
\end{figure*}

Now let us compare the numerical results for the pion form
factor obtained in
MIA (\ref{Fpi}) with that of the traditional IA.
Let us choose for
the comparison, for example, the null--component of the current.

To obtain the pion form factor in IA we proceed in the same
way as while obtaining
(\ref{Fc conv}) of the preceding Section.
Now, however,

1) the decomposition (\ref{j=Fc})
of the IA matrix current element over the state set
(\ref{p1m1p2m2}) is realized following (\ref{j=j1+j2}),

2) the parameterization of the one--particle matrix element
is given by (\ref{j=f1f2}), (\ref{Bal}) (instead of (\ref{j=f1})),

3) the CG coefficient (\ref{Klebsh}) in (\ref{wfp1p2})
are for pion quantum numbers.

Acting in the same way as while obtaining
(\ref{Fc conv}), and using the null--component
of the current matrix element, we can write
the pion form factor in IA in the following form:
$$
F_\pi(Q^2)
= \frac{M_\pi}{4}\frac{\sqrt{2\,(2\,M_\pi^2 + Q^2)}}
{4\,M_\pi^2 + Q^2}\,\frac{n_c}{\sqrt{1 + Q^2/4M^2}}
$$
$$
\times
\int\,\sqrt{\frac{s}{s'}}\,\frac{d\sqrt{s}\,d\sqrt{s'}}
{\sqrt{(s - 4\,M^2)(s' - 4\,M^2)}}
$$
$$
\times
\frac{(s + s' + Q^2)^3\,Q^2}{[\lambda(s\,,-Q^2\,,s')]^{3/2}}
\frac{1}{(s\,s')^{1/4}}
\frac{1}{\sqrt{s'}\,(s + Q^2)}\,\varphi(s)\, \varphi(s')
$$
$$
\times
\left\{(s + s' + Q^2)\left[G^u_E(Q^2) + G^{\bar d}_E(Q^2)\right]
\cos(\omega_1 + \omega_2)\right .
$$
\begin{equation}
+
\frac{1}{M}\,\xi(s,Q^2,s')\left.
\left[G^u_M(Q^2) + G^{\bar d}_M(Q^2) \right]\sin(\omega_1 + \omega_2)\right\}.
\label{Fpi conv}
\end{equation}
Here $M_\pi$ = 139.5702$\pm$0.0004 MeV \cite{PDG} is mass of pion.

The normalization condition $F_\pi(0) = 1$
is satisfied for the form factor (\ref{Fpi conv})
if the wave functions (\ref{phi(s)}) satisfy
(\ref{norm nc}).

To compare the numerical results given by the
Eqs.(\ref{Fpi}), (\ref{g_0})  with that given by
(\ref{Fpi conv}) let us calculate the pion form factor using the
wave function
(\ref{HO-wf}) with the parameters of the calculations presented
in Figs.3 and 4. The results are shown in the Fig.5.
The results obtained with the use of the parameterization
(\ref{Fc fin}), (\ref{g_0}) differ essentially from that
obtained without such parameterization (\ref{Fpi conv}).
The form factor calculated in our approach describes the
existing experimental data adequately.

\begin{figure*}
\centerline{\epsfxsize=0.4\textwidth \epsfbox{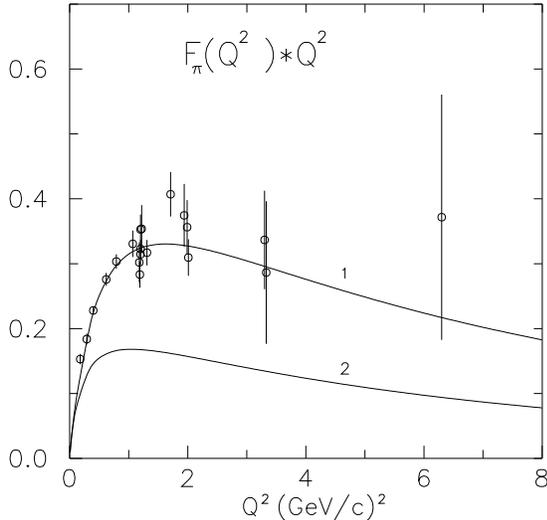}}
\caption{
$Q^2 F(Q^2)$ for MIA (1) and for IA (2). Results of calculation
with wave function (\protect\ref{HO-wf}). Parameters are the
same as in Fig.3.}
\end{figure*}

Let us emphasize once again that the form factor obtained in MIA
does not depend on the choice of coordinate frame. This is an
important advantage of our relativistic MIA.

\section{Conclusion}

Let us summarize the results.
\begin{enumerate}

\item
A new approach to the electromagnetic properties of
two--particle composite
systems is developed. The approach is based on IF RHD.

\item The main novel feature of this approach is the new method
of construction of the matrix element of the electroweak current
operator.
The electroweak current matrix element satisfies the
relativistic covariance conditions and in the case of the electromagnetic
current also the conservation law automatically.

\item The method of the construction of the current operator
matrix element consists of the extraction of the invariant part
-- the reduced matrix element on the Lorentz group (form
factor) -- and the covariant part defining the transformation
properties of the current. The form factors contain all the
dynamical information about transition. The properties of the
system as well as the approximations used are formulated in
terms of form factors, which in general have to be
considered as generalized functions.

\item The approach makes it possible to formulate relativistic
impulse approximation (modified impulse approximation --
MIA) in such  a way that the Lorentz--covariance of the current
is ensured. In the electromagnetic case the current conservation
law is ensured, too.

\item The results of the calculations are
unambiguous:  they do not depend on the choice of the coordinate
frame and on the choice of "good" components of the current as
it takes place in the standard form of light--front dynamics.
\item The formalism enables one to solve in part the problem of
connection of RHD and QFT by comparison of RHD with the
dispersion approach. In this paper RHD is compared with a
modified approach where dispersion--relation integrals over
composite--particle mass are used.

\item The effectiveness of
the approach is demonstrated by the calculation of the
electroweak structure of the pion. Our approach gives good
results for the pion electromagnetic form factor in the whole
range of momentum transfers available for experiments at present
time.

\end{enumerate}

\section*{Acknowledgements}
The authors thank V.V.Andreev for helpful discussions.
This work was supported in part by the Program "Russian
Universities -- Basic Researches" (grant No. 02.01.28).



\end{document}